\documentclass[a4paper,12pt]{article}
\usepackage{amssymb,graphicx,amsmath,array,verbatim,cite,multirow,arydshln,cases}

\def\p{\partial}
\newcommand{\vs}[1]{\vspace{#1 mm}}
\newcommand{\hs}[1]{\hspace{#1 mm}}
\newcommand{\bpm}{\begin{pmatrix}}
\newcommand{\epm}{\end{pmatrix}}
\renewcommand{\theequation}{\thesection.\arabic{equation}}

\newcommand{\R}{\mathbb{R}}
\newcommand{\C}{\mathbb{C}}
\newcommand{\Z}{\mathbb{Z}}

\newcommand{\tr}{{\rm Tr}}
\newcommand{\D}{\mathcal D}

\newcommand{\ba}{\left( \begin{array}}
\newcommand{\ea}{\end{array} \right)}
\newcommand{\be}{\begin{equation}}
\newcommand{\ee}{\end{equation}}
\newcommand{\beq}{\begin{eqnarray}}
\newcommand{\eeq}{\end{eqnarray}}
\newcommand{\beann}{\begin{eqnarray*}}
\newcommand{\eeann}{\end{eqnarray*}}

\newcommand{\diag}{{\rm diag}\,}

\makeatletter
\@addtoreset{equation}{section}
\def\theequation{\thesection.\arabic{equation}}
\makeatother

\begin{document}
\begin{titlepage}
\null
\begin{flushright}
TIT/HEP-604 \\
KUNS-2277 \\
July, 2010
\end{flushright}
\vskip 1.0cm
\begin{center}
  {\Large \bf Classification of BPS Objects \\
\vskip 0.3cm
  in $\mathcal{N} = 6$ Chern-Simons Matter Theory}
\vskip 2.0cm
\normalsize
\renewcommand\thefootnote{\alph{footnote}}

{\large
Toshiaki Fujimori$^{\dagger}$\footnote{fujimori(at)th.phys.titech.ac.jp},
Koh Iwasaki$^{\dagger \ddagger}$\footnote{iwasaki(at)th.phys.titech.ac.jp},
Yoshishige Kobayashi$^\dagger$\footnote{yosh(at)th.phys.titech.ac.jp }
and Shin Sasaki$^\dagger$\footnote{shin-s(at)th.phys.titech.ac.jp}
}
\vskip 0.5cm
  {\it
  $^\dagger$Department of Physics, Tokyo Institute of Technology \\
  Tokyo 152-8551, JAPAN \\
  \vskip 0.5cm
  $^\ddagger$Department of Physics, Kyoto University \\
  Kyoto 606-8502, JAPAN
}
\vskip 1.5cm
\begin{abstract}
We investigate BPS conditions
preserving $n/12 \ (n=1, \cdots, 6)$ supersymmetries
in the Aharony-Bergman-Jafferis-Maldacena (ABJM) model.
The BPS equations are classified
in terms of the number of preserved supercharges and
 remaining subgroups of the $SU(4)_R$ symmetry.
We study structures of a map between projection conditions
for the supercharges in eleven dimensions
and those in the ABJM model.
The BPS configurations in the ABJM model can be interpreted as
known BPS objects in eleven-dimensional M-theory,
such as intersecting M2, M5-branes, M-waves, KK-monopoles and M9-branes.
We also show that these BPS conditions reduce to
those in $\mathcal{N}=8$ super Yang-Mills theory
via the standard D2-reduction procedure
in a consistent way with the M-theory interpretation of the BPS conditions.
\vskip 0.5cm
\end{abstract}
\end{center}

\end{titlepage}

\newpage
\setcounter{footnote}{0}
\renewcommand\thefootnote{\arabic{footnote}}
\tableofcontents
\pagenumbering{arabic}
\section{Introduction}
\label{introduction}
In the past two years a great amount of effort has been invested
in a new class of superconformal field theories in three dimensions.
Especially two models have been extensively studied.
The first is the three-dimensional $\mathcal{N}=8$
superconformal field theory with an exotic three-algebra structure
proposed by Bagger, Lambert and Gustavsson (BLG model) \cite{BaLa, Gu}
and the other is the three-dimensional $\mathcal{N}=6$
superconformal Chern-Simons matter theory with $U(N)\times U(N)$ gauge symmetry proposed by Aharony-Bergman-Jafferis-Maldacena (ABJM model)
\cite{Aharony:2008ug}.
The ABJM model with Chern-Simons level $k$ is expected to
describe the low-energy effective theory of $N$ coincident M2-branes
located on the $\mathbb{C}^4/\mathbb{Z}_k$ orbifold fixed point.
One way to investigate this conjecture is
to study the BPS equations and
study whether they admit BPS objects that can be
identified with M-theoretical objects
such as M2 and M5-branes and so on.
Several configurations containing M2, M5 and other objects
have been found to be solutions of the classical equations
(BPS equations or equations of motion).
These can be found, for example, in
\cite{Hanaki:2008cu, Terashima:2008sy, Arai:2008kv, Fujimori:2008ga,
Kim:2009ny, Kawai:2009rc, Terashima:2009fy}.
However the complete catalog of solutions of the above equations
has not yet been obtained.

In the present paper,
we make further investigation on BPS configurations in the ABJM model.
Rather than finding explicit solutions of the BPS equations,
here we focus on the classification of BPS conditions and
the physical interpretations of the BPS objects.
Similar analysis has been carried out
both in the BLG and ABJM model \cite{Jeon:2008bx, Jeon:2008zj, Berman:2009}.
In this viewpoint we show that co-dimension one and two BPS objects
in ABJM model are systematically classified
in terms of the number of remaining supersymmetry (SUSY) and
the unbroken subgroup of the $SU(4)_R$ symmetry.
We classify the BPS conditions by using
a projection matrix $\mathcal A$ which specifies the preserved
supercharges via projection condition $\mathcal A \epsilon = \epsilon$
for supersymmetry transformation parameters $\epsilon$.
We also make an analysis on the physical interpretation of the BPS
conditions by finding a map between the conditions in eleven dimensions
and those in the ABJM model.
We show that BPS objects in the ABJM model can be interpreted
as M-theoretical objects such as M2-branes, M5-branes and so on.
We also consider the reduction of the BPS conditions to ten dimensions by the novel Higgs mechanism \cite{Mukhi:2008}.
We show that the BPS objects in the ABJM model are consistently reduced to
those in the $\mathcal{N}=8$ super Yang-Mills theory
which are identified with various BPS objects in type IIA string theory.

The organization of this paper is as follows.
In the next section, we briefly review the ABJM model and
introduce conventions, notations that we use throughout this paper.
In section 3, we classify the $1/2$ BPS equations of the ABJM model
in terms of supersymmetry projection matrices.  
In section 4, we discuss the supersymmetry projection conditions  
for the M2-branes and the $\mathbb{C}^4/\mathbb{Z}_k$ orbifold
from eleven-dimensional viewpoint.
After that we relate the M-theoretical objects
to the $1/2$ BPS states in the ABJM model.
In section 5, BPS configurations
with less than 6 preserved supercharges are discussed.
In section \ref{reduction}, we perform the reduction of the BPS equations
to the one in the $\mathcal{N}=8$ super Yang-Mills theory
through the Higgs mechanism and analyze the physical meaning of the solutions.
Section \ref{Conclusions} is conclusions and discussions.
Detailed expressions of the BPS equations are presented in Appendix.

\section{ABJM model} \label{models}
In this section, we briefly review the ABJM model and
introduce our notations and conventions that we use throughout this paper.
The ABJM model is a $(2+1)$-dimensional $\mathcal N = 6$
Chern-Simons matter system that
consists of $U(N) \times U(N)$ gauge fields $A_{\mu},
\hat{A}_{\mu}$ with Chern-Simons levels $k$ and $-k$,
four complex scalar fields $Y^A$ $(A=1,2,3,4)$ and their superpartners $\psi_A$
in the bifundamental representation of the gauge group\footnote{In this paper we basically employ the notations in \cite{Benna:2008zy}, but slightly modified them such that the Chern-Simons level $k$ appears as an overall coefficient of the action.}.
This model exhibits a manifest global $SU(4)_R \cong Spin(6)_R$ symmetry,
for which
$Y^A$ and $\psi_A$ are in the fundamental (upper indices) and
antifundamental (lower indices) representations.
This model is expected to describe
the low-energy effective world-volume theory of $N$ coincident M2-branes
probing $\C^4/\Z_k$ in eleven dimensions.
The bosonic part of the action is given as follows,
\beq
S &=& - \frac{k}{2\pi} \int d^3 x \, \tr \left[ D_\mu Y^A D^\mu Y^\dagger_A + \frac{2}{3} \Upsilon_A^{BC} (\Upsilon_A^{BC})^\dagger \right] \notag \\
&{}& + \frac{k}{4\pi} \int \! d^3 x \ \epsilon^{\mu \nu \rho} \mathrm{Tr} \left[ A_{\mu} \partial_{\nu} A_{\rho} + \frac{2i}{3} A_{\mu} A_{\nu} A_{\rho} - \hat{A}_{\mu} \partial_{\nu} \hat{A}_{\rho} - \frac{2i}{3} \hat{A}_{\mu} \hat{A}_{\nu} \hat{A}_{\rho} \right],
\eeq
where the covariant derivative is defined as
$D_{\mu} Y^A \equiv \p_{\mu} Y^A + i A_{\mu} Y^A - i Y^A \hat{A}_{\mu}$
and $\Upsilon_A^{BC}$ is given by
\beq
\Upsilon_A^{BC} ~\equiv~ Y^B Y_A^\dagger Y^C + \frac{1}{2} \delta_A^B \left( Y^C Y_D^\dagger Y^D - Y^D Y_D^\dagger Y^C \right) - ( B \leftrightarrow C ) \phantom{\frac12}.
\eeq
Hereafter, we will mainly discuss the supersymmetric variations of the fermions and BPS equations.

To obtain the BPS equations in the ABJM model,
we consider the supersymmetric variation of the fermionic fields.
In the notation with the manifest $SU(4)_R$ symmetry,
they are given by \cite{Gaiotto:2008cg, Terashima:2008sy}
\beq
\delta \psi_A = \left( \gamma^\mu D_\mu Y^B \delta_A^C + \Upsilon_A^{BC} \right) ( \Gamma_i )_{BC} \, \epsilon_i.
\label{eq:susy_variation}
\eeq
Here, $\epsilon_i~(i=1,\cdots,6)$ are (2+1)-dimensional Majorana spinor parameters which are in the vector representation $\mathbf 6$ of $SO(6)_R$ and $\Gamma^i \ (i = 1, \cdots, 6)$ are the matrices satisfying
\beq
\Gamma_i \Gamma_j^\dagger + \Gamma_j \Gamma_i^\dagger = 2 \delta_{ij}, \hs{10} (\Gamma_i)_{AB} = - (\Gamma_i)_{BA}, \hs{10} \frac{1}{2} \epsilon^{ABCD} (\Gamma_i)_{CD} = (\Gamma_i^\dagger)^{AB}. \eeq
Throughout this paper, we will use the following explicit forms
\beq
\begin{array}{cclccccccccc}
\Gamma_1 &=& \ba{cc} \sigma_2 &      \\  & \sigma_2 \ea, &\hs{10}&
\Gamma_2 &=& \ba{cc} - i \sigma_2 &  \\  & i \sigma_2 \ea, &\hs{10}&
\Gamma_3 &=& \ba{cc} & i \sigma_2    \\ i \sigma_2 & \ea,& \\
\Gamma_4 &=& \ba{cc} & i \sigma_1    \\ - i \sigma_1 & \ea, &\hs{10}&
\Gamma_5 &=& \ba{cc} & - i \sigma_3  \\ i \sigma_3 & \ea, &\hs{10}&
\Gamma_6 &=& \ba{cc} & - \mathbf 1_2 \\ \mathbf 1_2 & \ea, &
\end{array}
\eeq
where $\sigma_i~(i=1,2,3)$ are the Pauli matrices.
The $Spin(6)_R$ generators $\Sigma_{ij}~(i,j=1,\cdots,6)$
are given in terms of $\Gamma_i$ as
\beq
\Sigma_{ij} \equiv \frac{i}{4} (\Gamma_i^\dagger \Gamma_j - \Gamma_j^\dagger \Gamma_i ).
\eeq
For the $SO(2,1)$ gamma matrices $\gamma^\mu~(\mu=0,1,2)$, we will use
\beq
\gamma_0 = \ba{cc} 0 & 1 \\ -1 & 0 \ea, \hs{10} \gamma_1 = \ba{cc} 0 & 1 \\ 1 & 0 \ea, \hs{10} \gamma_2 = \ba{cc} 1 & 0 \\ 0 & -1 \ea.
\label{eq:gamma3d}
\eeq
In this basis, the Majorana spinor condition
for the spinor parameters $\epsilon_i$ is simply given by
\beq
\epsilon_i = \epsilon_i^\ast,
\eeq
where $\epsilon_i^\ast$ denotes the complex conjugate of $\epsilon_i$.

\section{Classification of 1/2 BPS equations} \label{sec:1/2BPS}
\subsection{Supersymmetry projection conditions}
In this section, we classify 1/2 BPS equations
which correspond to the most elementary
BPS objects in the ABJM model.
They are characterized by the preserved supercharges
determined by the projection condition of the form
\beq
\gamma \Xi_{ij} \epsilon_j = \epsilon_i, \hs{10} (i,j=1,2,\cdots,6).
\label{eq:projection}
\eeq
Here $\gamma$ is a 2-by-2 matrix acting on spinor indices and
$\Xi$ is a 6-by-6 matrix acting on $SO(6)_R$ vector indices.
Once the matrices $\gamma$ and $\Xi$ are given, BPS equations can be obtained
by requiring that the supersymmetric variations of the fermions \eqref{eq:susy_variation} vanish
for the spinor parameters $\epsilon_i$ specified by the condition \eqref{eq:projection}.

Now, let us classify the supersymmetry projection matrix
$\mathcal A \equiv \gamma \otimes \Xi$
acting on $(\mathbf 2, \mathbf 6)$ of $SO(2,1) \times SO(6)_R$
by imposing conditions which lead to half BPS equations.
We assume that $\mathcal A$ is traceless
\beq
\tr \, \mathcal A = 0 ,
\eeq
where the trace is taken over $SO(2,1) \times SO(6)_R$ indices.
We also assume that the square of $\mathcal A$ is the identity map
\beq
\mathcal A^2 ~=~ (\gamma \otimes \Xi)^2 ~=~ \mathbf 1_2 \otimes \mathbf 1_6.
\label{eq:cond_square}
\eeq
Furthermore, the matrix elements of $\gamma$ and $\Xi$ should be real
since the spinor parameters $\epsilon_i$ are Majorana spinors
satisfying $\epsilon_i^\ast = \epsilon_i$.
For an operator $\mathcal A$ satisfying these conditions,
there are generically $12 \times \frac{1}{2} = 6$
linearly independent solutions to the condition \eqref{eq:projection}.

First, let us classify the 2-by-2 real matrix $\gamma$,
which can be written as a linear combination of
the unit matrix and the gamma matrices
$\gamma = a \mathbf 1_2 + b^\mu \gamma_\mu$.
By using $SO(2,1)$ Lorentz transformations,
we can always fix the matrix $\gamma$ as
\beq
\gamma ~=~ \left\{ \begin{array}{ccl}
a \mathbf 1_2 + b \gamma_0 & \hs{5} & \mbox{if}~~b^\mu b_\mu = - b^2 \\
a \mathbf 1_2 + b \gamma_2 & \hs{5} & \mbox{if}~~b^\mu b_\mu = \phantom{-} b^2
\end{array} \right.,
\eeq
where $a$ and $b$ are real parameters.
From the condition \eqref{eq:cond_square},
one finds that possible matrices are
$\gamma = \mathbf 1_2$, $\gamma_0$ or $\gamma_2$
(up to Lorentz transformation).
However, we can show that BPS equations are trivial
for the unit matrix $\gamma = \mathbf 1_2$.
Therefore, there are essentially two possibilities
for the 2-by-2 matrix $\gamma$
\beq
\gamma = \gamma_0~\mbox{or}~\gamma_2.
\eeq
Next, let us classify the 6-by-6 real matrix $\Xi$.  
Because of the condition \eqref{eq:cond_square},
the matrix $\Xi$ should satisfy
\beq
\Xi^2 ~=~ \left\{ \begin{array}{rcl} -\mathbf 1_6 & \hs{5} & \mbox{if $\gamma = \gamma_0$} \\ \mathbf 1_6 & \hs{5} & \mbox{if $\gamma=\gamma_2$} \end{array} \right..
\eeq
If $\Xi$ is an anti-symmetric matrix, its square has negative eigenvalues.
Therefore, the 2-by-2 matrix $\gamma$ should be $\gamma_0$
for an anti-symmetric matrix $\Xi$.
In this case, the matrix $\Xi$ satisfying $\Xi^2 = - \mathbf 1_2$
can always be fixed in the following standard form by $SO(6)_R$ transformations
\beq
\Xi ~\rightarrow~ B ~\equiv~ \pm
\ba{c|c|c}
i\sigma_2 & & \\ \hline
& i \sigma_2 & \\ \hline
& & i \sigma_2
\ea.
\label{B_matrix}
\eeq
On the other hand, if $\Xi$ is a symmetric matrix,
$\gamma$ should be $\gamma_2$ since all the eigenvalues of $\Xi^2$ are positive.
Any symmetric matrix $\Xi$ satisfying $\Xi^2 = \mathbf 1_6$
can always be diagonalized by $SO(6)_R$ transformations as
\beq
\Xi ~\rightarrow~ C^{(m,n)} \equiv~ \ba{c|c} \mathbf 1_m & \\ \hline & -\mathbf 1_n \ea, \hs{10} m + n = 6.
\eeq
In summary, the possible 1/2 BPS projection matrices can be fixed
by using $SO(2,1) \times SO(6)_R$ transformation in the following forms
\begin{numcases}
{\mathcal{A} = }
\gamma_0 \otimes B & \label{half_BPS1} \\
\gamma_2 \otimes C^{(m,n)} & \label{half_BPS2}
\end{numcases}
Note that the second condition guarantees that the $\mathcal{N} = (m,n), \
(m + n = 6)$ supersymmetries are preserved in terms of $(1+1)$-dimensional
viewpoint ($x^{0,1}$-directions). Later, we will see that this condition is relaxed to the $m + n
\le 6$ case when we consider configurations with less than 1/2 supersymmetries.

\subsection{BPS equations}
Let us derive the BPS equations corresponding to
the 1/2 BPS projection matrices
which we have determined above.
First, we consider the case $\mathcal A = \gamma_0 \otimes B$.
For the spinor parameters satisfying
$\gamma_0 B_{ij} \epsilon_j = \epsilon_i$,
the supersymmetric variations of the fermions are given by
\beq
\delta \psi_A &=& \left[ \gamma^0 D_0 Y^B ( \Gamma_i )_{BA} + \Upsilon_A^{BC} ( \Gamma_i )_{BC} \right] \epsilon_i + (\gamma^1 D_1 Y^B + \gamma^2 D_2 Y^B) ( \Gamma_i )_{BA} \, \epsilon_i \\
&=& \left[ D_0 Y^B ( \Gamma_j )_{BA} B_{ji} + \Upsilon_A^{BC} ( \Gamma_i )_{BC} \right] \epsilon_i + \left[ D_1 Y^B (\Gamma_i)_{BA} + D_2 Y^B (\Gamma_j)_{BA} B_{ji} \right] \gamma_1 \epsilon_i. \notag
\eeq
Requiring $\delta \psi_A = 0$,
we obtain the following 1/2 BPS equations
\beq
0 &=& D_0 Y^B ( \Gamma_j )_{BA} B_{ji} + \Upsilon_A^{BC} ( \Gamma_i )_{BC}, \label{eq:BPS0-1} \\
0 &=& D_1 Y^B (\Gamma_i)_{BA} - D_2 Y^B (\Gamma_j)_{BA} B_{ji}.  \label{eq:BPS0-2}
\eeq
The more explicit forms of these equations are
given in Appendix \ref{appendix:BPSeqs}
(see \eqref{eq:M2_1/2_a}-\eqref{eq:M2_1/2_b}).
These equations are compatible with the equations of motion
provided that the following Gauss' law equations are satisfied
\beq
\begin{aligned}
0 =& \ \frac{1}{2} \epsilon^{\mu \nu \rho} F_{\mu \nu} + i \left[ Y^A (D^\rho Y^A)^\dagger - (D^\rho Y^A) Y^{A\dagger} \right], \label{eq:Gauss1} \\
0 =& \ \frac{1}{2} \epsilon^{\mu \nu \rho} \hat F_{\mu \nu} + i \left[ (D^\rho Y^A)^\dagger Y^A - Y^{A\dagger} (D^\rho Y^A) \right]. \label{eq:Gauss2}
\end{aligned}
\eeq
These BPS equations are equivalent to those discussed in \cite{Kim:2009ny}.
A solution to these equations contains co-dimension two vortex type objects
studied in \cite{Auzzi:2009es, Mohammed:2010eb, Arai:2008kv}.

Next, let us consider the case $\mathcal A = \gamma_2 \otimes C^{(m,n)}$. For the spinor parameters satisfying $\gamma_2 C_{ij}^{(m,n)} \epsilon_j = \epsilon_i$, the supersymmetric variations for the fermions are given by
\beq
\delta \psi_A &=& \left[ \gamma^2 D_2 Y^B ( \Gamma_i )_{BA} + \Upsilon_A^{BC} ( \Gamma_i )_{BC} \right] \epsilon_i + (\gamma^0 D_0 Y^B + \gamma^1 D_1 Y^B ) ( \Gamma_i )_{BA} \, \epsilon_i \\
&=& \left[ D_2 Y^B ( \Gamma_j )_{BA} C_{ji}^{(m,n)} + \Upsilon_A^{BC} ( \Gamma_i )_{BC} \right] \epsilon_i + \left[ D_0 Y^B (\Gamma_i)_{BA} - D_1 Y^B (\Gamma_j)_{BA} C_{ji}^{(m,n)} \right] \gamma_0 \epsilon_i. \notag
\eeq
From the condition $\delta \psi_A = 0$,
we obtain the following 1/2 BPS equations
\beq
0 &=& D_2 Y^B ( \Gamma_j )_{BA} C_{ji}^{(m,n)} + \Upsilon_A^{BC} ( \Gamma_i )_{BC}, \label{eq:BPS1-1} \\
0 &=& D_0 Y^B (\Gamma_i)_{BA} - D_1 Y^B (\Gamma_j)_{BA} C_{ji}^{(m,n)}. \label{eq:BPS1-2}
\eeq
The more explicit forms of these equations are
given in Appendix \ref{appendix:BPSeqs}.
In addition to these equations, the Gauss' law equations
\eqref{eq:Gauss1} should also be satisfied.
Examples of solution to these equations contain co-dimension one fuzzy funnels
and domain walls in the massless and massive ABJM models
which have been investigated in \cite{Terashima:2008sy, Hanaki:2008cu}.

It is worthwhile to note that the discussions on the BPS conditions in this paper can be generalized to
the massive ABJM model which keeps maximal supersymmetry \cite{Gomis:2008vc}.
This is achieved by the following replacement in the BPS conditions,
\begin{eqnarray}
\Upsilon_A {}^{BC} \to \Upsilon_A {}^{BC} + \frac{1}{2} M_A {}^{[B} Y^{C]}, \qquad
M_A {}^B = m \mathrm{diag} (1,1,-1-1),
\end{eqnarray}
where $m$ is a mass parameter.

\section{BPS conditions from eleven-dimensional viewpoint} \label{sec:BPS_10d}
\subsection{Supersymmetry preserved by M2-branes and orbifold}
In the previous section, we have classified
1/2 BPS equations in the ABJM model.
To understand the physical meaning of the BPS conditions
in the M2-brane world-volume theory,
it is useful to analyze a map between
BPS projection matrices in eleven-dimensional space-time
$\R^{(2,1)} \times \C^4 / \Z_k$ and their counterparts in the ABJM model.

First, let us specify the supersymmetry preserved
by
the orbifold in eleven dimensions and $N$ coincident M2-branes located at the orbifold
fixed point.
Let $x^M~(M=0,1,\cdots,10)$ be the eleven-dimensional space-time coordinates
and $y^A~(A=1,2,3,4)$ be the complex coordinates of $\C^4$ defined by
\beq
y^A = x^{2A+1} + i x^{2A+2}.
\eeq
The world-volume coordinate $x^{\mu}$ is chosen so that
$x^{\mu} \ (\mu =0,1,2)$ is identified with
the space-time coordinate $x^M (M=0,1,2)$
while the transverse directions $y^A$ are identified with
the bi-fundamental fields $Y^A$ living on the world-volume.
The $\mathbb{Z}_k$ orbifold action on the transverse coordinates is defined by
\beq
y^A \sim e^{2\pi i \frac{n}{k}} y^A, \qquad (n = 1,2,\cdots, k).
\label{orbifold_action}
\eeq

In the presence of $N$ M2-branes
extending along $x^{0,1,2}$-directions,
the preserved supercharges (16 SUSY) are specified by
\beq
\hat \Gamma_{012} \xi = \xi,
\label{eq:M2_condition}
\eeq
where $\xi$ is an eleven-dimensional
Majorana spinor parameter with 32 components and
$\hat \Gamma_M~(M=0,1,\cdots,10)$ are eleven-dimensional gamma matrices
satisfying $\{ \hat \Gamma_M, \hat \Gamma_N \} = 2 \eta_{MN}$.
The Lorentz symmetry $SO(10,1)$ is broken by the M2-branes
to $SO(2,1) \times SO(8)$, under which the spinor parameters
satisfying \eqref{eq:M2_condition} transform as $(\mathbf 2, \mathbf 8_c)$.

In addition to the condition (\ref{eq:M2_condition}),
the supersymmetric transformation parameter is  
further projected by the $\Z_k$ orbifold action.
The $\Z_k$ orbifold action on $\C^4$, which is a subgroup of $SO(8)$,
is defined by (\ref{orbifold_action}) and
the corresponding group elements acting on spinors are
\beq
\exp \left( \frac{2\pi n}{k} J \right) \in Spin(8), \hs{10} J \equiv i \sum_{A=1}^4 \hat \Sigma_{2A+1,2A+2}, \hs{10} (n = 1,\cdots,k).
\label{eq:orbifold_action}
\eeq
Here, $\hat \Sigma_{IJ}~(I,J=3,\cdots,10)$ are the $so(8)$ generators
in the spinor representation defined by
\beq
\hat \Sigma_{IJ} = \frac{i}{4} [ \hat \Gamma_I , \hat \Gamma_J ].
\eeq
The supercharges preserved in the orbifold are
those which are invariant under the orbifold action
\beq
\exp \left( \frac{2\pi n}{k} J \right) \xi = \xi, \hs{10} (n=1,\cdots,k).
\label{eq:orbifold}
\eeq
The orbifold
 projection breaks the $SO(8)$ symmetry
down to $SU(4) \times U(1)$ subgroup corresponding to
the generators which commutes with $J$.
This $SU(4)$ subgroup is nothing but
the $SU(4)_R$ symmetry in the ABJM model.
Explicitly, the generators of
$su(4) \oplus u(1)$ subalgebra are given by
\beq
S^A{}_B \equiv - \frac{1}{2} [ \hat \Gamma_+^A, \hat \Gamma_{-B} ],
\eeq
where $\hat \Gamma^A_+$ and $\hat \Gamma_{-A}~(A=1,2,3,4)$ are
raising and lowering operators
\beq
\hat \Gamma^A_+ \equiv \frac{\hat \Gamma_{2A+1} + i \hat \Gamma_{2A+2}}{2}, \hs{10} \hat \Gamma_{-A} \equiv \frac{\hat \Gamma_{2A+1} - i \hat \Gamma_{2A+2}}{2}.
\eeq
Under the breaking of the symmetry
$SO(8) \rightarrow SU(4) \times U(1)$,
the representation $\mathbf 8_c$ of $SO(8)$
decomposes into the following representations of $SU(4)$,
\beq
\mathbf 8_c \rightarrow \mathbf 6_0 \oplus \mathbf 1_2 \oplus \mathbf 1_{-2},
\eeq
where the subscript indicates the charge
under the $U(1)$ generator $-iJ = S^A{}_A$.
Therefore, the components of the spinor $\xi$
which are invariant under the orbifold action
form a basis of the vector space
$(\mathbf 2, \mathbf 6)$ of $SO(2,1) \times SU(4)$
\footnote{
Note that in the case of $k=1,2$
the $\mathbf 1_2 \oplus \mathbf 1_{-2}$ components
also satisfy Eq.\,\eqref{eq:orbifold},
so that $(\mathbf 2, \mathbf 8_c)$
is preserved by the orbifold action.
},
which is spanned by the following normalized spinors
\begin{eqnarray}
\begin{aligned}
&
| \alpha\,;\,\downarrow\,\downarrow\,\uparrow\,\uparrow\, \rangle,~~~~
| \alpha\,;\,\uparrow\,\uparrow\,\downarrow\,\downarrow\, \rangle,~~~~
| \alpha\,;\,\uparrow\,\downarrow\,\downarrow\,\uparrow\, \rangle,
\\
&
| \alpha\,;\,\downarrow\,\uparrow\,\uparrow\,\downarrow\, \rangle,~~~~
| \alpha\,;\,\downarrow\,\uparrow\,\downarrow\,\uparrow\, \rangle,~~~~
| \alpha\,;\,\uparrow\,\downarrow\,\uparrow\,\downarrow\, \rangle,\,
\label{orbifolded_spinor}
\end{aligned}
\end{eqnarray}
while $(\mathbf 2, \mathbf 1_2 \oplus \mathbf 1_{-2})$ part
is spanned by the following spinors
\beq
| \alpha\,;\,\uparrow\,\uparrow\,\uparrow\,\uparrow\, \rangle, \hs{10}
| \alpha\,;\,\downarrow\,\downarrow\,\downarrow\,\downarrow\, \rangle,
\eeq
where $\alpha = 1,2$ is an index of (2+1)-dimensional spinor
and each arrow indicates the eigenvalue $\pm \frac{1}{2}$ of
$\hat \Sigma_{2A+1,\,2A+2}$.
Namely, 16 supercharges specified by (\ref{eq:M2_condition}) is reduced
to 12 SUSY expressed by (\ref{orbifolded_spinor}). This is identified
with the $\mathcal{N} = 6$ supersymmetry in the ABJM model.
Throughout this paper, we will use the following
orthonormal basis for the vector space $(\mathbf 2, \mathbf 6)$
of $SO(2,1) \times SU(4)$
\beq
\psi^{(\alpha,1)}{} = \frac{| \alpha\,;\,\downarrow\,\downarrow\,\uparrow\,\uparrow\, \rangle + | \alpha\,;\,\uparrow\,\uparrow\,\downarrow\,\downarrow\, \rangle}{\sqrt{2}} ,~~~~
\psi^{(\alpha,2)}{} = i \frac{| \alpha\,;\,\downarrow\,\downarrow\,\uparrow\,\uparrow\, \rangle - | \alpha\,;\,\uparrow\,\uparrow\,\downarrow\,\downarrow\, \rangle}{\sqrt{2}} , \notag \\
\psi^{(\alpha,3)}{} = \frac{| \alpha\,;\,\uparrow\,\downarrow\,\downarrow\,\uparrow\, \rangle + | \alpha\,;\,\downarrow\,\uparrow\,\uparrow\,\downarrow\, \rangle}{\sqrt{2}} ,~~~~
\psi^{(\alpha,4)}{} = i \frac{| \alpha\,;\,\uparrow\,\downarrow\,\downarrow\,\uparrow\, \rangle - | \alpha\,;\,\downarrow\,\uparrow\,\uparrow\,\downarrow\, \rangle}{\sqrt{2}} , \\
\psi^{(\alpha,5)}{} = \frac{| \alpha\,;\,\downarrow\,\uparrow\,\downarrow\,\uparrow\, \rangle + | \alpha\,;\,\uparrow\,\downarrow\,\uparrow\,\downarrow\, \rangle}{\sqrt{2}} ,~~~~
\psi^{(\alpha,6)}{} = i \frac{| \alpha\,;\,\downarrow\,\uparrow\,\downarrow\,\uparrow\, \rangle - | \alpha\,;\,\uparrow\,\downarrow\,\uparrow\,\downarrow\, \rangle}{\sqrt{2}}. \notag
\eeq
For this choice of the basis of $(\mathbf 2, \mathbf 6)$,
the elements of $SO(6) \cong SU(4)/\Z_2$ are expressed by
orthogonal matrices $O^T O = \mathbf 1_6$.

The eleven-dimensional spinor parameter $\xi$
can be projected onto $(\mathbf 2, \mathbf 6)$
by using a projection operator $P$ (12-by-32 matrix)
which has the following structure
\beq
P^{(\alpha,i)}{}_{\hat \alpha} \equiv \left( \psi^{(\alpha,i)\dagger}{} \right)_{\hat \alpha},~~~~(\alpha = 1,2,~i=1,\cdots,6),~~~(\hat \alpha = 1,\cdots,32).
\eeq
where $\hat \alpha$ is the index of the eleven-dimensional spinor.
This projection operator satisfies
\beq
P P^\dagger = \mathbf 1_2 \otimes \mathbf 1_6, \hs{10}
P S^A{}_B = ( \mathbf 1_2 \otimes \mathcal S^A{}_B) P, \hs{10}
P J = 0,
\eeq
where $(\mathcal S^A{}_B)_{ij} \equiv i (\Sigma_{ij})^A{}_B$
are the generators of $su(4) \cong so(6)$
in the vector representation $\mathbf 6$.
The first equation follows from
the orthonormality of the basis $\{ \psi^{(\alpha,i)} \}$.
On the other hand, the second and the third equations, which indicate
the transformation property of $P$ under $SU(4)_R \times U(1)$,
follow from the fact that the spinors $\psi^{(\alpha,i)}$ are
in $\mathbf 6_0$ of $SU(4) \times U(1)$.

For later convenience, let us define a projection operator $\widetilde P$
(4-by-32 matrix) which projects the 32-component spinor
$\xi$ onto $(\mathbf 2, \mathbf 1_2 \oplus \mathbf 1_{-2})$
\beq
\widetilde P^{(\alpha,i)}{}_{\hat \alpha} \equiv \left( \tilde \psi^{(\alpha,i)\dagger}{} \right)_{\hat \alpha},~~~~(\alpha = 1,2,~i=1,2),~~~(\hat \alpha = 1,\cdots,32),
\eeq
where $\tilde \psi^{(\alpha,i)}_{\hat \alpha}$ are the basis of
$(\mathbf 2, \mathbf 1_2 \oplus \mathbf 1_{-2})$ given by
\beq
\tilde \psi^{(\alpha,1)} = \frac{\left| \alpha \,;\, \uparrow \, \uparrow \, \uparrow \, \uparrow \, \right>+\left| \alpha \,;\, \downarrow \, \downarrow \, \downarrow \, \downarrow \, \right>}{\sqrt{2}}, \hs{10}
\tilde \psi^{(\alpha,2)} = i \frac{\left| \alpha \,;\, \uparrow \, \uparrow \, \uparrow \, \uparrow \, \right> - \left| \alpha \,;\, \downarrow \, \downarrow \, \downarrow \, \downarrow \, \right>}{\sqrt{2}}.
\eeq
The projection operator $\widetilde P$ has the following properties
\beq
&\widetilde P P^\dagger = 0, \hs{10} \widetilde P \, \hat \Gamma_{012} = 0, \hs{10} \widetilde P \widetilde P^\dagger = \mathbf 1_2 \otimes \mathbf 1_2,& \\
&\widetilde P \left( S^A{}_B + \frac{i}{4} \delta^A{}_B J \right) = 0, \hs{10}
\widetilde P J = (\mathbf 1_2 \otimes 2 i \sigma_2) \widetilde P.&
\eeq
Since the four-dimensional linear space
$(\mathbf 2, \mathbf 1_2 \oplus \mathbf 1_{-2})$
is the orthogonal complement of
$(\mathbf 2, \mathbf 6_0)$ in $(\mathbf 2, \mathbf 8_c)$,
the following identity holds
\beq
P^\dagger P + \widetilde P^\dagger \widetilde P = \frac{\mathbf 1_{32} + \hat \Gamma_{012}}{2}.
\label{eq:id_projection}
\eeq

\subsection{1/2 BPS supersymmetry projection conditions from eleven dimensions}
So far, we have seen that the 12 supercharges preserved by the M2-branes
and orbifold are in $(\mathbf 2, \mathbf 6)$
of $SO(2,1) \times SU(4)$. They are specified by the conditions
\beq
\hat \Gamma_{012} \, \xi = \xi , \hs{10}
\exp \left( \frac{2\pi n}{k} J \right) \xi = \xi.
\label{eq:conditions}
\eeq
Here, the first condition has been imposed by the existence of
 M2-branes extending along $x^{0,1,2}$-directions
and the second condition is by the orbifold.
The supersymmetric transformation parameter
$\epsilon \in (\mathbf 2, \mathbf 6)$ in the ABJM model is given by
\beq
\epsilon = P \, \xi.
\eeq
In the following, we will observe that the BPS objects in
the ABJM model are interpreted as M-theory branes
in $\R^{(2,1)} \times \C^4 / \Z_k$ by analyzing
a map between projection matrices for eleven-dimensional SUSY parameters
and those in the ABJM model.

Let us consider a projection condition for the eleven-dimensional spinor
which are specified by the following equation
\beq
\hat \Gamma \, \xi = \xi.
\label{eq:condition2}
\eeq
Here, we assume that the operator $\hat \Gamma$
is invariant under the charge conjugation
since the spinor parameter $\xi$ is a Majorana spinor.
In addition, we assume that the operator $\hat \Gamma$ satisfies
\beq
[\hat \Gamma,\, \hat \Gamma_3 \hat \Gamma_4 \cdots \hat \Gamma_{10}] = 0,
\label{eq:SO(8)-chirality}
\eeq
since the spinor $\xi$ satisfying \eqref{eq:conditions}
has a definite chirality as an $SO(8)$ spinor.
Let us define maps $f$ and $\tilde{f}$ by
\beq
f : ~ \hat \Gamma ~ \longmapsto ~ \mathcal A \equiv P \, \hat \Gamma P^\dagger, \hs{10}
\tilde f : ~ \hat \Gamma ~ \longmapsto ~ \mathcal{\widetilde A} \equiv \widetilde P \, \hat \Gamma P^\dagger,
\label{operator_mapping}
\eeq
which map eleven-dimensional projection matrices $\hat \Gamma$
to those in the ABJM model. The operators
$\mathcal A : (\mathbf 2, \mathbf 6) \rightarrow (\mathbf 2, \mathbf 6)$ and
$\mathcal{\widetilde A} : (\mathbf 2, \mathbf 6) \rightarrow (\mathbf 2,
\mathbf 1 \oplus \mathbf 1)$ are 12-by-12 and 4-by-12 matrices respectively.
Using these operators, the equation \eqref{eq:condition2}
can be translated into the following conditions for the spinor parameters
$\epsilon = P \, \xi$,
\beq
\mathcal A \, \epsilon = \epsilon, \hs{10}
\mathcal{\widetilde A} \, \epsilon = 0,
\eeq
where we have used the condition \eqref{eq:id_projection}
and the fact that $\widetilde{P} \, \xi = 0$
for the spinor parameters satisfying \eqref{eq:conditions}.
Now, let us look for BPS projection matrices in eleven dimensions
which are mapped to the 1/2 BPS projection matrices in (2+1) dimensions
discussed in section \ref{sec:1/2BPS}.
Such matrices are determined from the condition
that the matrices (\ref{operator_mapping}) satisfy
\beq
\mathcal A^2 = \mathbf 1_2 \otimes \mathbf 1_6, \hs{10}
\tr \, \mathcal A = 0, \hs{10}
\mathcal{\widetilde A}=0.
\label{eq:red1/2cond}
\eeq
Note that the reality condition $\mathcal A^\ast = \mathcal A$ is
automatically satisfied since the operators $P$, $\widetilde P$ and $\hat \Gamma$
are invariant under the charge conjugation.
We assume that the operators $\hat \Gamma$ takes the form
\beq
\hat \Gamma = \frac{1}{p!} \omega_{M_1 M_2 \cdots M_p} \hat \Gamma_{M_1 M_2 \cdots M_p},
\eeq
where $\omega_{M_1 M_2 \cdots M_p}$ is a real $p$-form
in the eleven-dimensional space-time.
By using the relation $\hat \Gamma_{012} \, \xi = \hat \Gamma_{34\cdots10} \, \xi = \xi$,
one finds that independent operators are given by
\beq
\hat \Gamma ~=~ \left\{ \begin{array}{lcl}
\hat \Gamma^{(0)} &\equiv& \displaystyle \hat \Gamma_\mu \phantom{\bigg[} \\
\hat \Gamma^{(2)} &\equiv& \displaystyle \frac{1}{2} \omega_{IJ} \hat \Gamma_{\mu IJ} \phantom{\bigg[} \\
\hat \Gamma^{(4)} &\equiv& \displaystyle \frac{1}{4!} \omega_{IJKL} \hat \Gamma_{\mu IJKL} \phantom{\bigg[}
\end{array} \right.~~~~(\mu = 0,1,2, ~~~~I,J,K,L=3,\cdots,10).
\label{eq:hat_A}
\eeq
Note that the condition \eqref{eq:SO(8)-chirality}
is satisfied only when $\hat \Gamma$ contains
even number of $\Gamma^I~(I=3,\cdots,10)$.
For the operators of the form \eqref{eq:hat_A},
the corresponding operators $\mathcal A$ take the form
\beq
\mathcal A = \gamma_\mu \otimes \Xi,
\eeq
where $\gamma_\mu$ is the $(2+1)$-dimensional gamma matrix
given in $\eqref{eq:gamma3d}$ and
$\Xi$ is a 6-by-6 matrix acting on $\mathbf 6$ of $SO(6)$.
This form of the operator $\mathcal A$ implies that
the traceless condition $\tr \, \mathcal A = 0$ is
automatically satisfied for arbitrary $\Xi$.
Now, let us examine which types of operators satisfy
the conditions \eqref{eq:red1/2cond}
and what types of M-theoretical objects are described by
the corresponding BPS equations for each case of \eqref{eq:hat_A}.

\paragraph{$\bullet$ $\hat \Gamma = \hat \Gamma^{(0)}$ \\}
First, let us consider the case $\hat \Gamma = \hat \Gamma_\mu$.
Since the square of the operator
$\mathcal A = P \, \hat \Gamma_\mu P^\dagger$ is
\beq
\mathcal A^2 = P \, (\hat \Gamma_\mu)^2 P^\dagger,
\eeq
the gamma matrix $\hat \Gamma_\mu$
should be in the spacelike direction
because of the first condition in \eqref{eq:red1/2cond}.
Therefore, we can always set $\hat \Gamma_\mu = \hat \Gamma_2$
by using $SO(2,1)$ transformations without loss of generality.
Since the spinor parameter satisfies
$\hat \Gamma_{012} \xi = \hat \Gamma_{34\cdots10} \xi = \xi$,
the following three conditions are equivalent
\beq
\hat \Gamma_2 \xi = \xi \hs{5} \Longleftrightarrow \hs{5}
\hat \Gamma_{01} \xi = \xi \hs{5} \Longleftrightarrow \hs{5}
\hat \Gamma_{0134\cdots10} \xi = \xi.
\label{eq:wave_M9}
\eeq
The second condition
implies the existence of wave-type solutions, called the
gravitational Brinkman waves \cite{Brinkmann} or M-waves
which have momenta in the $x^1$-direction.
The third condition corresponds to a BPS object with $(9+1)$-dimensional
world-volume that is called M9-brane \cite{Bergshoeff:1998bs}.

Under the maps $f$ and $\tilde f$,
the operator $\hat \Gamma = \hat \Gamma_2$ reduces to
\beq
\mathcal A = \gamma_2 \otimes \mathbf 1_6, \hs{10}
\mathcal{\widetilde A} = 0.
\eeq
This is just the 1/2 BPS projection matrix
in (\ref{half_BPS2}) with $m=6,~n = 0$.
Therefore the M2-branes (we call this "fiducial M2-branes" on which we are considering the world-volume
theory),
M-waves and M9-branes can co-exist preserving half of 12 SUSY, see
Table \ref{tab:M2wave} for the configuration.
BPS objects that correspond to the projection $\mathcal A = \gamma_2
\otimes \mathbf 1_6$ contain co-dimension one solutions.
This fact is consistent with the intersection rules of various M-theoretical objects obtained in the supersymmetry algebra \cite{Bergshoeff:1997bh}.

\begin{table}[ht]
\begin{center}
\begin{tabular}{cccc:cc:cc:cc:cc}
& 0 & 1 & 2 & 3 & 4 & 5 & 6 & 7 & 8 & 9 & 10 \\
\mbox{M2} & $\bullet$ & $\bullet$ & $\bullet$ &  &  &&&&&& \\
\mbox{M9}
& $\bullet$ & $\bullet$ &  & $\bullet$ & $\bullet$ & $\bullet$ & $\bullet$ & $\bullet$ &$\bullet$ & $\bullet$ & $\bullet$ \\
\mbox{M-wave} & $\bullet$ & $\bullet$ &  &  &  &&&&&&
\end{tabular}
\caption{A possible configuration that corresponds
to the condition (\ref{eq:wave_M9}).
The black dot means that that directions are filled
by the corresponding objects.
Note that the dashed line indicates
the fact that the $\mathbb Z_k$ orbifold action acts on
the $(3,4),\, (5,6),\, (7,8),\,(9,10)$ planes separately.}
\label{tab:M2wave}
\end{center}
\end{table}

\paragraph{$\bullet$ $\hat \Gamma = \hat \Gamma^{(2)}$ \\}
Next, let us consider the case
$\hat \Gamma = \frac{1}{2} \omega_{IJ} \hat \Gamma_{\mu I J}$.
To find $\hat \Gamma$ for which
the conditions \eqref{eq:red1/2cond} are satisfied,
it is convenient to classify the anti-symmetric tensor
$\omega_{IJ}~(I,J=3,\cdots,10)$ in terms of representations of $SU(4)_R$.
The anti-symmetric tensor $\omega_{IJ}$ is in $\mathbf{28}$ of $SO(8)$,
which decomposes into the following
$SU(4) \times U(1) \cong SO(6) \times U(1)$ representations
\beq
\mathbf{28} ~\rightarrow~ \mathbf 6_2 \oplus (\mathbf{15}_0 \oplus \mathbf 1_0) \oplus \mathbf 6_{-2}.
\eeq
In this 28-dimensional linear space, the kernel of the map
$\tilde f : \hat \Gamma \mapsto \mathcal{\widetilde A}$ is given by
\beq
{\rm Ker} \, \tilde f = \mathbf{15}_0 \oplus \mathbf 1_0.
\eeq
Therefore, the third condition in \eqref{eq:red1/2cond}
is satisfied only when the 2-form $\omega_{IJ}$
belongs to $\mathbf{15}_0 \oplus \mathbf 1_0$.
Note that the representation $\mathbf{15}_0 \oplus \mathbf 1_0$
corresponds to $(1,1)$-type 2-forms $\omega_{A \bar B}$,
which are invariant under the orbifold action.
The first condition in \eqref{eq:red1/2cond} is
satisfied only when the gamma matrix
$\hat \Gamma_\mu$ is in the timelike direction.
Therefore, the matrix $\hat \Gamma = \hat \Gamma^{(2)}$ satisfying
the conditions \eqref{eq:red1/2cond}
can always be set in the following form
\beq
\hat \Gamma = \frac{1}{2} \omega_{IJ} \hat \Gamma_{0 IJ}, \hs{10} ( \omega_{IJ} dx^I \wedge dx^J = \omega_{A \bar B} dy^A \wedge d\bar y^B ).
\label{eq:A2}
\eeq
Since $\mathbf{15}$ is the adjoint representation of $SO(6)$,
the corresponding matrix $\hat \Gamma$ is mapped to
$\mathcal A$ whose 6-by-6 part $\Xi$ is an anti-symmetric matrix.
On the other hand, the singlet part $\mathbf 1$ does not affect
both $\mathcal A$ and $\mathcal{\widetilde A}$ since the singlet is
in the kernel of the map $f: \hat \Gamma \rightarrow \mathcal A$
\beq
{\rm Ker} \, f = \mathbf 6_2 \oplus \mathbf 1_0 \oplus \mathbf 6_{-2}.
\eeq
By appropriately choosing the singlet part
(trace part of $\omega_{A \bar B}$),
the 2-form $\omega_{A \bar B}$ becomes
a volume form of a complex plane $\C \subset \C^4$.
Since the projection condition \eqref{eq:A2} indicates that the BPS
object has $(2+1)$-dimensional world-volume, we conclude that
the operators of the form \eqref{eq:A2}
correspond to M2-branes stretching along complex planes
specified by the $(1,1)$-type 2-form $\omega_{A \bar B}$.

By using the fact that the spinor parameter satisfies
$\hat \Gamma_{012} \xi = \hat \Gamma_{34\cdots10} \xi = \xi$, we can show that the following two conditions are equivalent
\beq
\frac{1}{2} \omega_{IJ} \hat \Gamma_{0IJ} \xi = \xi \hs{5} \Longleftrightarrow \hs{5}
\frac{1}{6!} \tilde \omega_{IJKLMN} \hat \Gamma_{0IJKLMN} \xi = - \xi,
\label{eq:M2_KK}
\eeq
where $\tilde \omega \equiv \ast \omega$ is the Hodge dual of $\omega$
with respect to the metric on $\C^4/\Z_k$.
The second condition can be interpreted as KK monopoles
\cite{Gross:1983hb, Gueven:1992hh} extending along $(6+1)$-dimensional
world-volume determined by the volume form $\tilde{\omega}_{IJKLMN}$.
This fact is again an indication of co-existence of two intersecting
M2-branes and KK monopoles.

As an example, let us consider the matrix
$\hat \Gamma = \pm \hat{\Gamma}_0 \hat{\Gamma}_9 \hat{\Gamma}_{10}$
corresponding to M2-branes extending along $x^{0,9,10}$-directions.
In this case, the projected matrices
$\mathcal A$ and $\mathcal{\widetilde A}$ are given by
\beq
\mathcal A = \gamma_0 \otimes B, \qquad
\widetilde{\mathcal A} = 0.
\eeq
This is just the 1/2 BPS condition (\ref{half_BPS1}) in the ABJM model.
Therefore, the intersecting M2-branes with KK-monopoles are described by
the 1/2 BPS equations \eqref{eq:BPS0-1} and \eqref{eq:BPS0-2}
(see Table \ref{tab:M2M2} for a possible configuration).

\begin{table}[ht]
\begin{center}
\begin{tabular}{cccc:cc:cc:cc:cc}
& 0 & 1 & 2 & 3 & 4 & 5 & 6 & 7 & 8 & 9 & 10 \\
\mbox{M2} & $\bullet$ & $\bullet$ & $\bullet$ &  &  &&&&&& \\
\mbox{M2}
& $\bullet$ &  &  &  &  &  &  &  & & $\bullet$ & $\bullet$ \\
\mbox{KK} & $\bullet$ &  &  & $\bullet$ & $\bullet$ &$\bullet$&$\bullet$&$\bullet$&$\bullet$&&
\end{tabular}
\caption{Intersecting M2-branes and KK-monopoles.}
\label{tab:M2M2}
\end{center}
\end{table}

\paragraph{$\bullet$ $\hat \Gamma = \hat \Gamma^{(4)}$ \\}
Next, we consider the case
$\hat \Gamma = \frac{1}{4!} \omega_{IJKL} \Gamma_{\mu I J K L}$.
As in the previous case, let us classify the 4-form
in terms of the representation of $SU(4)_R$.
The 4-form $\omega_{IJKL}$ is
in $\mathbf{35} \oplus \mathbf{35}'$ of $SO(8)$,
which decomposes into the following $SU(4) \times U(1)$ representations
\beq
\mathbf{35} \oplus \mathbf{35}' \rightarrow \mathbf 1_4 \oplus (\mathbf 6_2 \oplus \mathbf{10}_2) \oplus ( \mathbf 1_0 \oplus \mathbf{15}_0 \oplus \mathbf{20}_0 ) \oplus (\mathbf 6_{-2} \oplus \overline{\mathbf{10}}_{-2} ) \oplus \mathbf 1_{-4}.
\eeq
The kernel of the map
$\tilde f : \hat \Gamma \mapsto \mathcal{\widetilde A}$ is given by
\beq
{\rm Ker} \, \tilde f &=& \mathbf 1_4 \oplus \mathbf{10}_2 \oplus ( \mathbf 1_0 \oplus \mathbf{15}_0 \oplus \mathbf{20}_0 ) \oplus \overline{\mathbf{10}}_{-2} \oplus \mathbf 1_{-4}.
\eeq
The third condition in \eqref{eq:red1/2cond} is satisfied
only when the 4-form $\omega_{IJKL}$ is in ${\rm Ker} \, \tilde f$.
The first condition in \eqref{eq:red1/2cond} is
satisfied only when the gamma matrix
$\hat \Gamma_\mu$ is in the spacelike direction.
Therefore, the matrix $\hat \Gamma = \hat \Gamma^{(4)}$ satisfying
the conditions \eqref{eq:red1/2cond}
can always be set in the following form
\beq
\hat \Gamma^{(4)} = \frac{1}{4!} \omega_{IJKL} \hat \Gamma_{2IJKL}, \hs{10}
\omega \not \in \mathbf 6_2 \oplus \mathbf 6_{-2}.
\label{eq:A4}
\eeq
The kernel of the map
$f : \hat \Gamma \mapsto \mathcal A$ is given by
\beq
{\rm Ker} \, f &=& \mathbf 1_4 \oplus (\mathbf 6_2 \oplus \mathbf{10}_2) \oplus \mathbf{15}_0 \oplus (\mathbf 6_{-2} \oplus \overline{\mathbf{10}}_{-2}) \oplus \mathbf 1_{-4}.
\eeq
The matrix $\hat{\Gamma}$ should be in $\mathrm{Ker} \tilde{f}$
to be a 1/2 BPS projection matrix in the ABJM model and
components in $\mathrm{Ker} f$ are irrelevant for the matrix $\mathcal A$.
This means that the matrix $\mathcal A$ is determined by $
\mathbf 1_0 \oplus \mathbf{20}_0$ components,
which correspond to the symmetric tensor representation of $SO(6)$.
Therefore, 1/2 BPS objects specified by the operator \eqref{eq:A4}
are described in the ABJM model
by the 1/2 BPS equations with symmetric matrices $\Xi$,
which can be fixed to $C^{(m,n)}$ by $SO(6)$ transformations as we have
seen in section \ref{sec:1/2BPS}.

Let $\omega^{(m,n)} = \frac{1}{4!} \omega_{IJKL}^{(m,n)} d x^I \wedge d x^J \wedge d x^K \wedge d x^L$ be the 4-forms corresponding to
the matrices $\Xi = C^{(m,n)}$.
Assuming that $\omega^{(m,n)}$ is invariant under
the $Spin(m) \times Spin(n) \subset SU(4)$ transformations,
we can find the following examples\footnote{
Note that there exists an ambiguity since we can add elements of ${\rm Ker} \, f \cap {\rm Ker} \, \tilde f$ without changing the reduced operators $\mathcal A$ and $\mathcal{\widetilde A}$.
}
of 4-forms
\beq
\omega^{(6,0)} &=& \frac{1}{16} d y^A \wedge d y^B \wedge d \bar y_A \wedge d \bar y_B, \\
\omega^{(5,1)} &=& \frac{1}{64} \big[ \mathcal J_{AB} \mathcal J_{\bar C \bar D} - \delta_{A \bar C} \delta_{B \bar D} + \delta_{A \bar D} \delta_{B \bar C} \big] d y^A \wedge d y^B \wedge d \bar y^{\bar C} \wedge d \bar y^{\bar D},
\label{eq:51_form} \\
\omega^{(4,2)} &=& \frac{1}{4} d y^1 \wedge d y^3 \wedge d \bar y^1 \wedge d \bar y^3, \\
\omega^{(3,3)} &=& {\rm Re} \, d \check y^1 \wedge {\rm Re} \, d \check y^2 \wedge {\rm Re} \, d \check y^3 \wedge {\rm Re} \, d \check y^4, \phantom{\frac{1}{2}}
\eeq
where $\mathcal J \equiv \Gamma_6$ is
a $Spin(5) \cong USp(4)$ invariant tensor and
$\check y^A$ are coordinates rotated by $\exp \left( \frac{\pi i}{2} \Sigma_{16}\right) \in SU(4)_R$
\beq
\check y^A \equiv \left( e^{\frac{\pi i}{2} \Sigma_{16}} \cdot y \right)^A .
\eeq
The other 4-forms $\omega^{(2,4)},\,\omega^{(1,5)},\,\omega^{(0,6)}$ can be obtained by flipping the signs of $\omega^{(4,2)},\,\omega^{(5,1)},\,\omega^{(6,0)}$ and using $SU(4)$ transformations.
Since the spinor parameter satisfies $\hat{\Gamma}_{012} \xi =
\xi$, the condition $\omega_{IJKL} \hat \Gamma_{2 I J K L} \xi = \xi$
can be rewritten as
\beq
\omega_{IJKL} \hat \Gamma_{0 1 I J K L} \xi = \xi.
\label{M5_operator}
\eeq
The BPS objects corresponding to the operators
(\ref{M5_operator}) have $(5+1)$-dimensional world-volume and
are identified with M5-branes that share one world-volume direction
with our fiducial M2-branes.
If the 4-form $\omega$ is a calibration,
it determines a calibrated submanifold of $\C^4/\Z_k$.
A related work in the BLG model can be found in \cite{Krishnan:2008zm}.

The 4-form $\omega^{(6,0)}$ is proportional to
the square of a K\"ahler form on $\C^4/\Z_k$,
whose calibrated submanifold is
the four-dimensional complex submanifold of $\C^4/\Z_k$.
Therefore, in addition to the M-waves and M9-branes
given in Table \ref{tab:M2wave},
the 1/2 BPS equation with $m=6,\,n=0$ admits calibrated M5-branes.
For $\omega^{(4,2)}$ and $\omega^{(3,3)}$,
the operator $\hat \Gamma$ is given by
\beq
\hat \Gamma ~=~ \frac{1}{4!} \omega_{IJKL} \hat \Gamma_{2IJKL} ~=~
\left\{ \begin{array}{lc} \hat{\Gamma}_{23478} & \mbox{for $\omega = \omega^{(4,2)}$} \\ \hat{\Gamma}_{2 \check 3 \check 5 \check 7 \check 9} & \mbox{for $\omega = \omega^{(3,3)}$} \end{array} \right..
\eeq
Therefore, the BPS equations \eqref{eq:BPS1-1} and \eqref{eq:BPS1-2}
with $\mathcal N = (4,2)$ and $(3,3)$ describe the configurations of the fiducial M2-branes
ending on M5-branes (see Table \ref{tab:M2M5(4,2)} and \ref{tab:M2M5(3,3)}).
Note that because of the conditions
$\hat \Gamma_{012} \xi = \hat \Gamma_{34\cdots10} \xi = \xi$, the following two conditions are equivalent
\beq
\frac{1}{4!} \omega_{IJKL} \hat \Gamma_{2IJKL} \xi = \xi \hs{5} \Longleftrightarrow \hs{5}
\frac{1}{4!} \tilde \omega_{IJKL} \hat \Gamma_{2IJKL} \xi = \xi,
\label{eq:M5_M5}
\eeq
where $\tilde \omega \equiv \ast \omega$ is the Hodge dual of $\omega$
with respect to the metric on $\C^4/\Z_k$.
\begin{table}[ht]
\begin{center}
\begin{tabular}{cccc:cc:cc:cc:cc}
& 0 & 1 & 2 &  3 & 4 & 5 & 6 & 7 & 8 & 9 & 10 \\
\mbox{M2} & $\bullet$ & $\bullet$ & $\bullet$ &  &  &&&&&& \\
\mbox{M5}
& $\bullet$ & $\bullet$ &  & $\bullet$ & $\bullet$ &  &  & $\bullet$ &$\bullet$ & &  \\
\mbox{M5} & $\bullet$ & $\bullet$ &  &  &  &$\bullet$&$\bullet$&&&$\bullet$& $\bullet$
\end{tabular}
\caption{M2-M5 configuration : $(m,n)=(4,2)$.}
\label{tab:M2M5(4,2)}
\end{center}
\end{table}

\vs{10}

\begin{table}[ht]
\begin{center}
\begin{tabular}{cccc:cc:cc:cc:cc}
& 0 & 1 & 2 & 3 & 4 & 5 & 6 & 7 & 8 & 9 & 10 \\
\mbox{M2} & $\bullet$ & $\bullet$ & $\bullet$ &  &  &&&&&& \\
\mbox{M5}
& $\bullet$ & $\bullet$ &  & $\bullet$ &  & $\bullet$  &  & $\bullet$ & & $\bullet$ &  \\
\mbox{M5} & $\bullet$ & $\bullet$ &  &  & $\bullet$ &&$\bullet$&&$\bullet$ && $\bullet$
\end{tabular}
\caption{M2-M5 configuration : $(m,n)=(3,3)$.}
\label{tab:M2M5(3,3)}
\end{center}
\end{table}

\section{BPS equations with less supersymmetries} \label{sec:BPS}
In the previous section, we have studied 1/2 BPS conditions that
preserve 6 SUSY among 12 SUSY in the ABJM model.
Once we consider more general configurations of M-theory branes,
it is possible to partially break the 6 SUSY down up to 1 SUSY.
In this section, we investigate BPS configurations
with less than 6 supersymmetries. 
In the first half of the following subsections,
we consider M2 and M5-branes that intersect with
our fiducial M2-branes where the M2 and M5-branes have
non-trivial angles with the orbifolded planes. Such a classification of M-branes in terms of
angles are deeply investigated in \cite{ohta:1998}.  
In the latter half, we discuss more general configurations
with multiple kinds of M2, M5-branes
which correspond to more than one projection conditions.

\subsection{M2-M2 intersections with angles}
In this subsection, we consider M2-branes which intersect with our
fiducial M2-branes at a point. The fiducial M2-branes are extending
along $x^{0,1,2}$-directions while the other M2-branes spans in the
two-dimensional subspace of $\mathbb{C}^4/\mathbb{Z}_k$
which is transverse to the $x^{0,1,2}$-directions.

Let $v^I_1,\,v^I_2~(I=3,4,\cdots,10)$ be two linearly independent vectors
indicating a plane in $\R^8$ along which M2-branes are extending.
The corresponding 1/2 BPS projection matrix in eleven dimensions is
\beq
\hat \Gamma = \frac{1}{2} (v_1^I v_2^J - v_2^I v_1^J) \Gamma_{0 I J}.
\label{eq:genericM2}
\eeq
Here, the normalization of the vectors should be
determined from the condition $\hat \Gamma^2 = \mathbf 1$.
This operator $\hat \Gamma$ is invariant up to normalization under
$GL(2,\R)$ transformations which mix the vectors $v_1$ and $v_2$.
Therefore, M2-branes are specified by points on the Grassmannian $G(2,\R^8)$,
which is described by 8-by-2 matrices with the following equivalence relation
\beq
\ba{cc}
v_1^3 & v_2^3 \\
v_1^4 & v_2^4 \\
\vdots & \vdots \\
v_1^{10} & v_2^{10}
\ea
\sim
\ba{cc}
v_1^3 & v_2^3 \\
v_1^4 & v_2^4 \\
\vdots & \vdots \\
v_1^{10} & v_2^{10}
\ea
g,
\hs{10} \forall g \in GL(2,\R).
\eeq
However, instead of real 8-by-2 matrix,
it is convenient to use the following complex 4-by-2 matrix
to make the $SU(4)_R$ symmetry manifest
\beq
\Lambda_{\rm M2} ~\equiv~
\ba{cc}
u_1^1 & u_2^1 \\
u_1^2 & u_2^2 \\
u_1^3 & u_2^3 \\
u_1^4 & u_2^4 \\
\ea,
\label{eq:LambdaM2}
\eeq
where $u^A_1$ and $u^A_2$ are complex vectors corresponding to $v_1$ and $v_2$
\beq
u^A_1 = v_1^{2A+1} + i v_1^{2A+2}, \hs{10}
u^A_2 = v_2^{2A+1} + i v_2^{2A+2}.
\eeq
The $SU(4)_R$ transformation acts on $\Lambda_{\rm M2}$ from the left
and $GL(2,\R)$ matrices acts from the right
$\Lambda_{\rm M2} \sim \Lambda_{\rm M2} g$.
Once the matrix $\Lambda_{\rm M2}$ is given
the vectors $v_1^I,\,v_2^I$ and the operator \eqref{eq:genericM2}
can be uniquely determined.
Without loss of generality,
we can always fix $\Lambda_{\rm M2}$ in the following form by using
$SU(4) \times U(1)$ and $GL(2,\R)$ transformations
\beq
\Lambda_{\rm M2} =
\ba{cc}
0 & 0 \\
0 & 0 \\
i \sin \theta & 0 \\
\cos \theta & i
\ea.
\eeq
For this form of $\Lambda_{\rm M2}$,
the operator \eqref{eq:genericM2} is given by
\beq
\hat \Gamma = \hat \Gamma_0 ( \sin \theta \, \hat \Gamma_8 + \cos \theta \, \hat \Gamma_9 ) \hat \Gamma_{10}.
\label{eq:4SUSY_11d}
\eeq
Under the maps $f$ and $\tilde f$,
the operator $\hat \Gamma$ reduces to
$\mathcal A = \gamma_0 \otimes \Xi$ and
$\mathcal{\widetilde A} = \gamma_0 \otimes \widetilde \Xi$ with
\beq
\Xi = - g^T \,\diag ( \, i \sigma_2 \, , \, i \sigma_2 \, , \, i \cos \theta \sigma_2 \, ) \, g, \hs{7}
\widetilde \Xi =
\ba{cccccc}
0 & 0 & 0 & 0 & -\sin \theta & 0 \\
0 & 0 & 0 & 0 & 0 & \sin \theta
\ea g,
\eeq
where $g$ is an $SO(6)$ matrix.
For generic values of the angle parameter $\theta$,
there are 4 components of the spinor parameters $\epsilon$ satisfying
\beq
\mathcal A \, \epsilon = \epsilon, \hs{10} \mathcal{\widetilde A} \, \epsilon = 0.
\label{eq:4SUSY}
\eeq
The first condition reduces 12 SUSY down to 4 SUSY which is consistent with the second condition.
As a result, 4 SUSY among 12 SUSY is preserved by the projection
conditions (\ref{eq:4SUSY}).
Therefore, M2-branes with generic values of the angle parameter $\theta$
are described by 1/3 BPS equations in the ABJM model\footnote{
Note that the M2-branes themselves are orthogonal with each other and, as one can see in Table \ref{tab:M2}, $\theta$ parameterizes
the angle between the second (not the fiducial) M2-branes and the hyperplanes on which the orbifold projection acts, see (\ref{orbifold_action}).
}.
The corresponding configuration is summarized in Table \ref{tab:M2}.
On the other hand, the supersymmetry enhances to 6 SUSY
(1/2 BPS) at $\theta = 0$ and $\theta = \pi$
since the conditions \eqref{eq:red1/2cond} are satisfied
and the M2-branes do not extend across the line between two different orbifolded planes.

\begin{table}[ht]
\begin{center}
\begin{tabular}{cccc:cc:cc:cc:cc}
& 0 & 1 & 2 & 3 & 4 & 5 & 6 & 7 & 8 & 9 & 10 \\
\mbox{M2} & $\bullet$ & $\bullet$ & $\bullet$ &  &  &&&&&& \\
$\mbox{M2}$
& $\bullet$ &  &  &  &  &  & & & $\underset{\sin \theta}{\circ}$ & $\underset{\cos \theta}{\circ}$ & $\bullet$  
\end{tabular}
\caption{1/3 BPS configuration of intersecting M2-branes.}
\label{tab:M2}
\end{center}
\end{table}

\subsection{M2-M5 intersections with angles}
Next, let us consider M5-branes extending along
$x^1$-direction and an arbitrary four-dimensional subspace in $\R^8$
spanned by four linearly independent vectors $v_a^I~(a=1,2,3,4)$.
The corresponding operator in eleven dimensions is
\beq
\hat \Gamma \propto \frac{1}{4!} \epsilon^{abcd} v_a^I v_b^J v_c^K v_d^L \hat \Gamma_{0 1 I J K L}.
\eeq
The operators of this form are specified by
points on the Grassmannian $G(4,\R^8)$.
As in the previous case, let us define complex vectors
$u_a^A \equiv v_a^{2A+1} + i v_a^{2A+2}~(a=1,2,3,4)$
and the 4-by-4 complex matrix $\Lambda_{\rm M5}$ by
\beq
\Lambda_{\rm M5}
\equiv
\ba{cccc}
u_1^1 & u_2^1 & u_3^1 & u_4^1 \\
u_1^2 & u_2^2 & u_3^2 & u_4^2 \\
u_1^3 & u_2^3 & u_3^3 & u_4^3 \\
u_1^4 & u_2^4 & u_3^4 & u_4^4
\ea,
\label{eq:LambdaM5}
\eeq
where $SU(4)$ transformations act from the left and
$GL(4,\R)$ transformations act from the right
$\Lambda_{\rm M5} \sim \Lambda_{\rm M5} g$.
By using the $SU(4)$ and $GL(4,\R)$ transformations,
the matrix $\Lambda_{\rm M5}$ can always be fixed as
\beq
\Lambda_{M5}
=
\ba{cccc}
1 & i \cos \theta_1 & 0 & 0 \\
0 & \sin \theta_1 & 0 & 0 \\
0 & 0 & 1 & i \cos \theta_2 \\
0 & 0 & 0 & \sin \theta_2 \ea.
\eeq
For this matrix $\Lambda_{\rm M5}$,
the operator $\hat \Gamma$ is mapped by $f$ and $\tilde{f}$ to
$\mathcal A = \gamma_0 \otimes \Xi$ and $\mathcal{\widetilde A} = \gamma_0 \otimes \widetilde \Xi$
with
\beq
\Xi &=& g^T \, \diag \big( \,1\, , \,1\, , \, \cos (\theta_1 - \theta_2) \,, \, \cos (\theta_1 + \theta_2) \, ,-1,-1 \, \big)\, g, \phantom{\bigg(}\\
\widetilde \Xi &=&
\ba{cccccc}
0 & 0 & -\sin(\theta_1-\theta_2) & 0 & 0 & 0 \\
0 & 0 & 0 & \sin(\theta_1+\theta_2) & 0 & 0 \ea \, g,
\eeq
where $g$ is an $SO(6)$ matrix.
For generic values of the angle parameters $\theta_1$ and $\theta_2$,
the conditions $\mathcal A \, \epsilon = \epsilon$ and
$\mathcal{\widetilde A} \, \epsilon = 0$ are satisfied
by $4$ components of the spinor parameters $\epsilon$.
Namely, 4 SUSY among 12 SUSY is preserved (1/3 BPS).
See Table \ref{tab:M2M5(2,2)} for the configuration.
If either of the angle parameters $\theta_1 \pm \theta_2$
becomes $0$ or $\pi$, the supersymmetry enhances to 5 SUSY (5/12 BPS).
If both of the angle parameters satisfy $\theta_1 \pm \theta_2 = 0$ or $\pi$,
the preserved supersymmetry becomes 6 SUSY (1/2 BPS)
and the corresponding configurations are those in
Table \ref{tab:M2M5(4,2)} $(\theta_1=0,\,\theta_2=0)$
or Table \ref{tab:M2M5(3,3)} $(\theta_1=\frac{\pi}{2},\,\theta_2=\frac{\pi}{2})$.

\begin{table}[ht]
\begin{center}
\begin{tabular}{cccc:cc:cc:cc:cc}
& 0 & 1 & 2 & 3 & 4 & 5 & 6 & 7 & 8 & 9 & 10 \\
\mbox{M2} & $\bullet$ & $\bullet$ & $\bullet$ &  &  &&&&&& \\
$\mbox{M5}$
& $\bullet$ & $\bullet$ &  & $\bullet$ & $\underset{\cos \theta_1}{\circ}$ & $\underset{\sin \theta_1}{\circ}$   &  & $\bullet$  & $\underset{\cos \theta_2}{\diamond}$ & $\underset{\sin \theta_2}{\diamond}$ &
\end{tabular}
\caption{1/3 BPS configuration of M2/M5-branes.}
\label{tab:M2M5(2,2)}
\end{center}
\end{table}

\subsection{M2-M2 intersections with less supersymmetries}
So far, we have considered only single sets of conditions
$\mathcal A \, \epsilon = \epsilon,~\mathcal{\widetilde A} \, \epsilon = 0$
corresponding to a single type of BPS objects.
If several sets of conditions are imposed on the spinor parameter $\epsilon$,
we can obtain BPS equations which admit multiple types of BPS objects.
First, let us classify BPS configurations consisting of M2-branes only.
Let $\Xi^{(I,J)}$ be 6-by-6 matrices
corresponding to M2-branes
extending along $x^{I,J}$-directions by
\beq
\gamma_0 \otimes \Xi^{(I,J)} = P \, \hat \Gamma_{0IJ} P^\dagger.
\eeq
We first study a condition that preserve 4 supercharges.
To find such a condition, we consider two sets of M2-branes
extending along $x^{5,6}$ and $x^{9,10}$-directions.
The corresponding projection matrices $\hat{\Gamma}_{056}, \hat{\Gamma}_{09(10)}$ are mapped to the one in ABJM model leading to the
following conditions
\beq
(-\gamma_0 \otimes \Xi^{(5,6)}) \epsilon = \epsilon, \hs{10}
(\gamma_0 \otimes \Xi^{(9,10)}) \epsilon = \epsilon,
\label{eq:M2M2cond2}
\eeq
where $\Xi^{(5,6)}$ and $\Xi^{(9,10)}$ are given by
\beq
\Xi^{(5,6)} = - \diag (-i \sigma_2,\, -i \sigma_2 ,\, i \sigma_2), \hs{10}
\Xi^{(9,10)} = - \diag (i \sigma_2 ,\, i \sigma_2 ,\, i \sigma_2).
\eeq
Clearly, the above conditions force the 8 SUSY among 12 SUSY to vanish resulting 4 remaining SUSY.
The conditions \eqref{eq:M2M2cond2} correspond to
an anti-M2-branes extending along $x^{5,6}$-directions and
an M2-branes extending along $x^{9,10}$-directions
(Table \ref{tab:intersectiong M2})
\footnote{
Here the overline on the "M2" means that the corresponding eleven-dimensional projector has extra minus sign
compared with the "M2" without overline which is defined through the projector of the form $\Gamma_{0IJ}$. We call this anti-M2-branes.
In the following, we use the overline to denote this interpretation.
}.
\begin{table}[ht]
\begin{center}
\begin{tabular}{cccc:cc:cc:cc:cc}
& 0 & 1 & 2 & 3 & 4 & 5 & 6 & 7 & 8 & 9 & 10 \\
$\mbox{M2}$ & $\bullet$ & $\bullet$ & $\bullet$ &  &  &&&&&& \\
$\overline{\mbox{M2}}$
& $\bullet$ &  &  &  &  & $\bullet$  & $\bullet$ &   &  &  &  \\
$\mbox{M2}$ & $\bullet$ &&  &  &  &&&&  & $\bullet$ & $\bullet$
\end{tabular}
\caption{1/3 BPS intersecting M2-branes.}
\label{tab:intersectiong M2}
\end{center}
\end{table}
In addition, arbitrary M2-branes can be added
without breaking further supersymmetry
if the corresponding matrix $\Xi$ takes the following form
\beq
\Xi = -\diag (i \sigma_2 ,\, i \sigma_2 ,\, i c \, \sigma_2 ),
\label{eq:1/3A}
\eeq
where $c$ is an arbitrary real parameter.
Let us see this fact in more detail.
Consider M2-branes extending
along a plane spanned by two vectors $v_1$ and $v_2$, then
we find that the corresponding BPS projection matrix
$\hat \Gamma = v_{[1}^I v_{2]}^J \hat \Gamma_{0IJ}$
reduces to $\Xi$ of the form \eqref{eq:1/3A}
if the matrix $\Lambda_{\rm M2}$ defined in \eqref{eq:LambdaM2} is given by
\beq
\Lambda_{\rm M2} =
\ba{cccc}
0 & 0 \\
a & -ia \\
0 & 0 \\
b & i b
\ea,
\eeq
where $a$ and $b$ are complex parameters satisfying
$|a|^2 + |b|^2 = 1$ and $|b|^2 - |a|^2 = c$.
This form of the matrix $\Lambda_{\rm M2}$ indicates
a three-dimensional submanifold of
the Grassmannian $G(2,\R^8)$ corresponding to
a family of two-dimensional planes in $\R^8$.
Note that the condition $\mathcal{\widetilde A} \, \epsilon = 0$
is automatically satisfied for the spinor parameter $\epsilon$
satisfying \eqref{eq:M2M2cond2} since $\mathcal{\widetilde A}$ takes the form
\beq
\mathcal{\widetilde A} = 2 \gamma_0 \otimes \ba{cccccc} 0 & 0 & 0 & 0 & {\rm Im}(ab) & \phantom{-} {\rm Re}(ab) \\ 0 & 0 & 0 & 0 & {\rm Re}(ab) & - {\rm Im}(ab) \ea.
\eeq
Therefore, the 1/3 BPS equation of this type admits
three parameter family of M2-branes.

The 1/6 BPS equations can be obtained
by adding the following supersymmetry projection condition
to \eqref{eq:M2M2cond2}
\beq
(\gamma_0 \otimes \Xi^{(7,8)}) \epsilon = \epsilon, \hs{10}
\Xi^{(7,8)} = - \diag ( \, i \sigma_2 ,\, -i \sigma_2 ,\, -i \sigma_2).
\eeq
This condition corresponds to M2-branes extending along $x^{7,8}$-directions.
In this case, we can add arbitrary M2-branes whose matrix $\Xi$ has the following form
\renewcommand\arraystretch{1.5}
\beq
\Xi = - \ba{c|c} i \sigma_2 & \\ \hline  & ~ \mbox{{\Huge $\ast$}}_4 \ea,
\label{eq:1/6A}
\eeq
where $\ast_4$ denotes arbitrary entries.
For example, we can add anti-M2-branes in $x^{3,4}$-directions
since the corresponding projection condition
takes the form
\renewcommand\arraystretch{1}
\beq
(- \gamma_0 \otimes \Xi^{(3,4)}) \epsilon = \epsilon, \qquad
\Xi^{(3,4)} = - \diag (-i \sigma_2,\, i \sigma_2 ,\, -i \sigma_2).
\eeq
An example of the 1/6 BPS intersecting M2-branes
is given in Table \ref{tab:M2_M2_anti-M2}.
\begin{table}[ht]
\begin{center}
\begin{tabular}{cccc:cc:cc:cc:cc}
& 0 & 1 & 2 & 3 & 4 & 5 & 6 & 7 & 8 & 9 & 10 \\
\mbox{M2} & $\bullet$ & $\bullet$ & $\bullet$ &  &  &&&&&& \\
$\overline{\mbox{M2}}$
& $\bullet$ &  &  & $\bullet$ & $\bullet$ &   &  &   &  &  &  \\
$\overline{\mbox{M2}}$
& $\bullet$ &  &  &  &  & $\bullet$  & $\bullet$ &   &  &  &  \\
\mbox{M2} & $\bullet$ &&  &  &  &&& $\bullet$  & $\bullet$  & &  \\
\mbox{M2} & $\bullet$ &&  &  &  &&&&  & $\bullet$ & $\bullet$
\end{tabular}
\caption{1/6 BPS intersecting M2-branes.}
\label{tab:M2_M2_anti-M2}
\end{center}
\end{table}
In general, the BPS projection matrix
$\hat \Gamma = v_{[1}^I v_{2]}^J \hat \Gamma_{0IJ}$
reduces to the matrix of the form \eqref{eq:1/6A}
if the matrix $\Lambda_{\rm M2}$ is given by
\beq
\Lambda_{\rm M2}
=
\ba{c|c}
U_1 & \\ \hline
 & U_2
\ea
\ba{cccc}
0 & 0 \\
a & -ia \\
0 & 0 \\
b & i b
\ea,
\eeq
where $|a|^2 + |b|^2 = 1$ and $U_1,\,U_2 \in SU(2)$.
Therefore, compared with the 1/3 BPS equation,
the 1/6 BPS equation of this type admits
larger class of M2-branes rotated by $SU(2) \times SU(2) \subset SU(4)_R$.
It is worthwhile to note that the M2-branes can be accompanied by KK-monopoles
through the Hodge dualized projection conditions of the spinor parameters.

\subsection{M2-M5 intersections with less supersymmetries}
Let us next consider M5-branes
which share the $x^1$-direction with our fiducial M2-branes.
The BPS configurations are classified
by the chirality (eigenvalues of $\gamma_2$)
of the preserved supercharges.
In a BPS configuration with
$\mathcal N = (m,n)$ preserved supercharges,
the M5-branes are characterized by the condition
$\gamma_2 \Xi_{ij} \epsilon_j = \epsilon_i$
with the symmetric matrix $\Xi$ of the form
\renewcommand\arraystretch{1.5}
\beq
\Xi = \ba{c|c|c}
\mathbf 1_m & & \\ \hline
& -\mathbf 1_n & \\ \hline
& & \, \mbox{{\Huge $\ast$}$_{6-m-n}$} \ea, \hs{10} m + n \leq 6,
\label{eq:(m,n)}
\eeq
where $\ast_{6-m-n}$ is an arbitrary entry.
Now, let us determine the matrix $\Lambda_{\rm M5}$
(defined in \eqref{eq:LambdaM5})
corresponding to $\Xi$ of the form \eqref{eq:(m,n)}.
Since arbitrary symmetric matrix $\Xi$ can be diagonalized
by $SO(6)_R$ transformations, it is sufficient to determine
a matrix $\Lambda_{\rm M5}$ corresponding to
the diagonal matrix of the form
\renewcommand\arraystretch{1}
\beq
\Xi = \diag( \, \overbrace{ 1,\phantom{(} \cdots \ , 1}^m,\, \overbrace{-1, \phantom{(} \cdots \ ,-1}^n, \, \overbrace{\ast, \phantom{(} \cdots \ ,\ast}^{6-m-n} \, ).
\eeq
The generic forms of $\Lambda_{\rm M5}$ can be obtained
by using $Spin(m) \times Spin(n) \times Spin(6-m-n)$ transformations
corresponding to $SO(m) \times SO(n) \times SO(6-m-n)$ transformations
which leave the form of \eqref{eq:(m,n)} unchanged.

First, we note the fact that any 4-by-4 matrix $\Lambda_{\rm M5}$
can be fixed in the following form by using
the $GL(4,\R)$ and $SU(4)_R$ transformations
\beq
\Lambda_{\rm M5}
=
\ba{cccc}
\sin \theta_1  & -i \sin \theta_1 & 0 & 0 \\
0 & 0 & \cos \theta_2 & i \cos \theta_2 \\
\cos \theta_1  & i \cos \theta_1 & 0 & 0 \\
0  & 0 & \sin \theta_2 & -i \sin \theta_2
\ea.
\label{eq:vec}
\eeq
For this matrix $\Lambda_{\rm M5}$, the matrix $\hat \Gamma$ reduces to
$\mathcal A = \gamma_2 \otimes \Xi$ with
\beq
\Xi = \diag \Big( \, 1 \,,\, 1 \,, -1 \,, -1 \,,\cos[2(\theta_1+\theta_2)],\cos[2(\theta_1-\theta_2)] \Big).
\label{eq:diag}
\eeq
Therefore, the matrix $\Xi$ for arbitrary M5-branes
extending along a four-dimensional plane in $\mathbb C^4/\Z_k$
can be obtained from \eqref{eq:diag} by using $SO(6) \cong SU(4)_R$ rotations.
If $SO(6)$ is restricted to $SO(5)$
generated by $\Sigma_{ij}~(i,j=2,\cdots,6)$,
the matrix $\Xi$ takes the form
\renewcommand\arraystretch{1.5}
\beq
\Xi = \ba{c|c} 1 & \\ \hline  & ~ \mbox{{\Huge $\ast$}}_5 \ea.
\label{eq:(1,0)}
\eeq
For this form of the matrix, the BPS configurations preserve
$\mathcal N = (m,n) = (1,0)$ supercharges (1/12 BPS) and
the corresponding matrix $\Lambda_{\rm M5}$ can be obtained from
\eqref{eq:vec} by using $Spin(5) \cong USp(4)$ transformations defined by
\renewcommand\arraystretch{1}
\beq
U^T \Gamma_1 U = \Gamma_1, \hs{10} U^\dagger U = \mathbf 1_4.
\eeq
Therefore, the 1/12 BPS equation preserving $\mathcal N = (1,0)$ SUSY
admits a family of M5-branes specified by
\beq
\Lambda_{\rm M5}
=
U^\dagger
\ba{cccc}
\sin \theta_1  & -i \sin \theta_1 & 0 & 0 \\
0 & 0 & \cos \theta_2 & i \cos \theta_2 \\
\cos \theta_1  & i \cos \theta_1 & 0 & 0 \\
0  & 0 & \sin \theta_2 &-i \sin \theta_2
\ea U, \hs{10} U \in USp(4).
\eeq
If we choose a specific matrix $U$ and angles $(\theta_1,\theta_2)$, we find that $\mathcal N = (1,0)$ SUSY admits
a configuration given in Table \ref{tab:(1,0)}.

\begin{table}[ht]
\begin{center}
\begin{tabular}{cccc:cc:cc:cc:cc}
& 0 & 1 & 2 & 3 & 4 & 5 & 6 & 7 & 8 & 9 & 10 \\
\mbox{M2} & $\bullet$ & $\bullet$ & $\bullet$ &  &  &&&&&& \\
$\mbox{M5}$
& $\bullet$ & $\bullet$ &  & $\bullet$ &  & $\bullet$ &  &  & $\bullet$ &  & $\bullet$ \\
$\overline{\mbox{M5}}$
& $\bullet$ & $\bullet$ &  & $\bullet$ &  &  & $\bullet$ & $\bullet$ &  &  & $\bullet$ \\
$\overline{\mbox{M5}}$
& $\bullet$ & $\bullet$ &  &  & $\bullet$ & $\bullet$ &  & $\bullet$ &  &  & $\bullet$ \\
$\overline{\mbox{M5}}$
& $\bullet$ & $\bullet$ &  &  & $\bullet$ &  & $\bullet$ &  & $\bullet$ &  & $\bullet$ \\
$\mbox{M5}$
& $\bullet$ & $\bullet$ &  & $\bullet$ & $\bullet$ &  &  &  &  & $\bullet$ & $\bullet$ \\
$\mbox{M5}$
& $\bullet$ & $\bullet$ &  &  &  & $\bullet$ & $\bullet$ &  &  & $\bullet$ & $\bullet$ \\
$\overline{\mbox{M5}}$
& $\bullet$ & $\bullet$ &  &  &  &  &  & $\bullet$ & $\bullet$ & $\bullet$ & $\bullet$
\end{tabular}
\caption{An example of $\mathcal N = (1,0)$ BPS configuration. The Hodge-dual branes are omitted.}
\label{tab:(1,0)}
\end{center}
\end{table}

In the above, we have used $Spin(5) \cong USp(4)$
to obtain the most generic form of the matrix $\Lambda_{\rm M5}$
which specify the M5-branes in $\mathcal N = (1,0)$ configurations.
If we restrict the transformation matrix $U$
to $Spin(4) \subset Spin(5)$ defined by
\beq
U^T \Gamma_i U = \Gamma_i,~~~~~(i=1,2),
\eeq
the matrix $\Xi$ is rotated by $SO(4)$ generated by
$\Sigma_{ij}~(i,j=3,\cdots,6)$ and takes the form
\renewcommand\arraystretch{1.5}
\beq
\Xi = \ba{c|c} \mathbf 1_2 & \\ \hline & ~ \mbox{{\Huge $\ast$}}_4 \ea.
\label{eq:(2,0)}
\eeq
Therefore, the most general form of the vectors for M5-branes
in $\mathcal N = (2,0)$ configurations can be obtained
from \eqref{eq:vec} by using
$Spin(4) \cong SU(2) \times SU(2)$.
An example of $\mathcal{N} = (2,0)$ configuration
is given in Table \ref{tab:(2,0)}.
\renewcommand\arraystretch{1}
\begin{table}[ht]
\begin{center}
\begin{tabular}{cccc:cc:cc:cc:cc}
& 0 & 1 & 2 & 3 & 4 & 5 & 6 & 7 & 8 & 9 & 10 \\
\mbox{M2} & $\bullet$ & $\bullet$ & $\bullet$ &  &  &&&&&& \\
$\mbox{M5}$
& $\bullet$ & $\bullet$ &  & $\bullet$ & $\bullet$ &  &  &  &  & $\bullet$ & $\bullet$ \\
$\mbox{M5}$
& $\bullet$ & $\bullet$ &  &  &  & $\bullet$ & $\bullet$ &  &  & $\bullet$ & $\bullet$ \\
$\overline{\mbox{M5}}$
& $\bullet$ & $\bullet$ &  &  &  &  &  & $\bullet$ & $\bullet$ & $\bullet$ & $\bullet$
\end{tabular}
\caption{$\mathcal N = (2,0)$ BPS configuration. The Hodge-dual branes are omitted.}
\label{tab:(2,0)}
\end{center}
\end{table}

To specify the form of the matrix $\Lambda_{\rm M5}$ corresponding to
$\mathcal N = (1,1)$ configurations, we have to first determine
a diagonal matrix of the form
\beq
\Xi = \diag ( 1 , -1, \ast , \ast , \ast , \ast ).
\eeq
This form of the matrix $\Xi$ can be
obtained from \eqref{eq:diag} by using
$\exp \left( i \frac{\pi}{2} \Sigma_{13} \right) \in SO(6)$,
which corresponds to the $SU(4)$ element given by
\beq
U' \equiv \frac{1}{\sqrt{2}}
\ba{c|c}
\mathbf 1_2 & \mathbf 1_2 \\ \hline
- \mathbf 1_2 & \mathbf 1_2
\ea.
\eeq
Therefore, the most general form of the vectors for M5-branes
in $\mathcal N = (1,1)$ configurations can be obtained
from \eqref{eq:vec} by acting $U'$ and
then using $Spin(4) \cong SU(2) \times SU(2)$ transformations.
An example of $\mathcal N = (1,1)$ configuration
is given in Table \ref{tab:(1,1)}.

\begin{table}[ht]
\begin{center}
\begin{tabular}{cccc:cc:cc:cc:cc}
& 0 & 1 & 2 & 3 & 4 & 5 & 6 & 7 & 8 & 9 & 10 \\
\mbox{M2} & $\bullet$ & $\bullet$ & $\bullet$ &  &  &&&&&& \\
$\mbox{M5}$
& $\bullet$ & $\bullet$ &  & $\bullet$ &  & $\bullet$ &  &  & $\bullet$ &  & $\bullet$ \\
$\overline{\mbox{M5}}$
& $\bullet$ & $\bullet$ &  & $\bullet$ &  &  & $\bullet$ & $\bullet$ &  &  & $\bullet$ \\
$\overline{\mbox{M5}}$
& $\bullet$ & $\bullet$ &  &  & $\bullet$ & $\bullet$ &  & $\bullet$ &  &  & $\bullet$ \\
$\overline{\mbox{M5}}$
& $\bullet$ & $\bullet$ &  &  & $\bullet$ &  & $\bullet$ &  & $\bullet$ &  & $\bullet$
\end{tabular}
\caption{$\mathcal N = (1,1)$ BPS configuration. The Hodge-dual branes are omitted. }
\label{tab:(1,1)}
\end{center}
\end{table}

By similar discussions, we can determine
the matrix $\Lambda_{\rm M5}$ for $\mathcal N = (m,n)$ configurations.
Note that the BPS equations considered here admits not only M5-branes
extending along a four-dimensional plane in $\C^4 / \Z_k$
but also other types of BPS objects.
For example, the BPS equations with $\mathcal N = (m,0)$
admits the wave-type and its Hodge dual M9-brane configurations.
The complete list of BPS equations for $\mathcal N = (m,n)$
are given in Appendix \ref{appendix:BPSeqs}.

\subsection{M2-branes ending on intersecting M5-branes}
In the previous subsection, we have analyzed M2-M5 configurations
in which all branes share one common direction ($x^1$-direction).
Here, let us consider BPS configurations of M2-branes
ending on M5-branes extending along both $x^{1,2}$-directions.
For such configurations,
the condition for the preserved supercharges are given by
\beq
\gamma_2 \Xi_{ij}  \epsilon_j = \epsilon_i, \hs{10}
\gamma_1 \Xi_{ij}' \epsilon_j = \epsilon_i,
\label{eq:condM5M5}
\eeq
where $\Xi$ and $\Xi'$ are 6-by-6 symmetric real matrix.
The most supercharges are preserved
when $[\gamma_2 \Xi, \gamma_1 \Xi' ] = 0$ and $\Xi^2 = \Xi'^2 = \mathbf 1_6$.
Up to $SU(4)_R$ transformations,
the matrices $\Xi$ and $\Xi'$ satisfying these conditions are given by
\beq
\Xi = \diag(\sigma_3, \sigma_3, \sigma_3), \hs{10}
\Xi' = \diag( \sigma_1 , \sigma_1 , \sigma_1 ).
\eeq
For these matrices, the condition \eqref{eq:condM5M5}
is satisfied by three components of the spinor parameter $\epsilon$.
Therefore the projections with these matrices are 1/4 BPS conditions.
The corresponding matrices $\Lambda_{\rm M5}$ and $\Lambda_{\rm M5}'$
for the matrices $\Xi$ and $\Xi'$ are given by
\beq
\Lambda_{\rm M5}
~=~
\ba{cccc}
1 & 0 & 0 & 0 \\
0 & i & 0 & 0 \\
0 & 0 & i & 0 \\
0 & 0 & 0 & -1
\ea, \hs{10}
\Lambda_{\rm M5}'
~=~
\ba{cccc}
1 & 0 & 0 & 0 \\
0 & i & 0 & 0 \\
0 & 0 & i & 0 \\
0 & 0 & 0 & i
\ea.
\label{eq:M5M5vec}
\eeq
In this BPS configuration, the M5-branes have
three-dimensional common world-volume ($x^{3,6,8}$-directions).
Obviously, we can also add objects specified by
\beq
\gamma_0 \Xi''_{ij} \epsilon_j = \epsilon_i, \hs{10}
\Xi'' \equiv \Xi \Xi' = \diag (i\sigma_2, i\sigma_2, i\sigma_2 ).
\eeq
This implies that M2-branes which intersect our fiducial M2-branes at a point
can also be added to the 1/4 BPS configuration. An example of the 1/4 BPS configuration is given in Table \ref{tab:1/4}.
\begin{table}[ht]
\begin{center}
\begin{tabular}{cccc:cc:cc:cc:cc}
& 0 & 1 & 2 & 3 & 4 & 5 & 6 & 7 & 8 & 9 & 10 \\
$\mbox{M2}$ & $\bullet$ & $\bullet$ & $\bullet$ &  &  &&&&&& \\
$\overline{\mbox{M2}}$
& $\bullet$ &  &  &  &  &   &  &  &  & $\bullet$ & $\bullet$ \\
$\mbox{M5}$ & $\bullet$ & $\bullet$ &  & $\bullet$ &  & & $\bullet$ &   & $\bullet$ &  & $\bullet$ \\
$\overline{\mbox{M5}}$ & $\bullet$ &  & $\bullet$ &  & $\bullet$ & $\bullet$ &  & $\bullet$ &  &  & $\bullet$
\end{tabular}
\caption{1/4 BPS configuration. The Hodge-dual branes are omitted.}
\label{tab:1/4}
\end{center}
\end{table}
There are also BPS configurations with two preserved supercharges (1/6 BPS),
for which the matrices $\Xi$, $\Xi'$ and $\Xi''$ are given by
\beq
\Xi = \diag ( \sigma_3, \sigma_3, \mbox{{\Large $\ast$}}_2 ), \hs{10}
\Xi' = \diag( \sigma_1, \sigma_1, \mbox{{\Large $\ast$}}_2 ), \hs{10}
\Xi'' = \diag( i\sigma_2, i\sigma_2, \mbox{{\Large $\ast$}}_2 ).
\eeq
These matrices correspond to M5-branes and M2-branes
and th matrices $\Lambda_{\rm M5}$ and $\Lambda_{\rm M2}$ for them
can be determined in a similar way as in the previous sections.
\begin{table}[ht]
\begin{center}
\begin{tabular}{cccc:cc:cc:cc:cc}
& 0 & 1 & 2 & 3 & 4 & 5 & 6 & 7 & 8 & 9 & 10 \\
$\mbox{M2}$ & $\bullet$ & $\bullet$ & $\bullet$ &  &  &&&&&& \\
$\mbox{M2}$
& $\bullet$ &  &  &  &  & $\bullet$ & $\bullet$ &  &  &  & \\
$\overline{\mbox{M2}}$
& $\bullet$ &  &  &  &  &   &  &  &  & $\bullet$ & $\bullet$ \\
$\mbox{M5}$ & $\bullet$ & $\bullet$ &  & $\bullet$ &  & & $\bullet$ &   & $\bullet$ &  & $\bullet$  \\
$\overline{\mbox{M5}}$ & $\bullet$ & $\bullet$ &  &  & $\bullet$ &  & $\bullet$ & $\bullet$ &  &  & $\bullet$  \\
$\mbox{M5}$ & $\bullet$ &  & $\bullet$ & $\bullet$ &  & $\bullet$ &  &  &  $\bullet$ &  & $\bullet$ \\
$\overline{\mbox{M5}}$ & $\bullet$ &  & $\bullet$ &  & $\bullet$ & $\bullet$ &  & $\bullet$ &  &  & $\bullet$
\end{tabular}
\caption{1/6 BPS configuration. The Hodge-dual branes are omitted.}
\label{tab:1/6}
\end{center}
\end{table}
The most generic BPS configurations in this class
preserve one supercharge (1/12 BPS) specified by
\renewcommand\arraystretch{1.5}
\beq
\Xi = \ba{c|c} \sigma_3 & \\ \hline  & \hspace{1mm} \mbox{{\Huge $\ast$}}_4 \ea, \hs{10}
\Xi' = \ba{c|c} \sigma_1 & \\ \hline  & \hspace{1mm} \mbox{{\Huge $\ast$}}_4 \ea.
\eeq
An example of 1/6 and 1/12 BPS configuration is given in Table \ref{tab:1/6} and \ref{tab:1/12}.
The BPS equations for these matrices are summarized in Appendix \ref{appendix:BPSeqs}.
\renewcommand\arraystretch{1.0}

\begin{table}[ht]
\begin{center}
\begin{tabular}{cccc:cc:cc:cc:cc}
& 0 & 1 & 2 & 3 & 4 & 5 & 6 & 7 & 8 & 9 & 10 \\
$\mbox{M2}$ & $\bullet$ & $\bullet$ & $\bullet$ &  &  &&&&&& \\
$\mbox{M2}$
& $\bullet$ &  &  & $\bullet$ & $\bullet$ &  &  &  &  &  & \\
$\mbox{M2}$
& $\bullet$ &  &  &  &  & $\bullet$ & $\bullet$ &  &  &  & \\
$\overline{\mbox{M2}}$
& $\bullet$ &  &  &  &  &   &  &  $\bullet$ & $\bullet$  &  & \\
$\overline{\mbox{M2}}$
& $\bullet$ &  &  &  &  &   &  &  &  & $\bullet$ & $\bullet$ \\
$\mbox{M5}$ & $\bullet$ & $\bullet$ &  & $\bullet$ &  & $\bullet$ &  & $\bullet$ &  &  & $\bullet$  \\
$\mbox{M5}$ & $\bullet$ & $\bullet$ &  &  & $\bullet$ & $\bullet$ &  &  & $\bullet$ &  & $\bullet$ \\
$\mbox{M5}$ & $\bullet$ & $\bullet$ &  & $\bullet$ &  & & $\bullet$ &   & $\bullet$ &  & $\bullet$  \\
$\overline{\mbox{M5}}$ & $\bullet$ & $\bullet$ &  &  & $\bullet$ & & $\bullet$ &$\bullet$ &  &  & $\bullet$ \\
$\mbox{M5}$ & $\bullet$ &  & $\bullet$ & $\bullet$ &  & $\bullet$ &  &  &  $\bullet$ &  & $\bullet$ \\
$\overline{\mbox{M5}}$ & $\bullet$ &  & $\bullet$ &  & $\bullet$ & $\bullet$ &  & $\bullet$ &  &  & $\bullet$ \\
$\overline{\mbox{M5}}$ & $\bullet$ &  & $\bullet$ & $\bullet$ &  &  & $\bullet$ & $\bullet$ &  &  & $\bullet$ \\
$\overline{\mbox{M5}}$ & $\bullet$ &  & $\bullet$ &  & $\bullet$ &  & $\bullet$ &  & $\bullet$ &  & $\bullet$
\end{tabular}
\caption{1/12 BPS configuration. The Hodge-dual branes are omitted.}
\label{tab:1/12}
\end{center}
\end{table}

\section{Reduction to $\mathcal{N} = 8$ super Yang-Mills theory \label{reduction}}
In this section,
we study the ten-dimensional interpretation of our BPS conditions
by reducing the ABJM model to the multiple D2-brane effective theory,
namely $(2+1)$-dimensional $\mathcal{N} = 8$ super Yang-Mills theory.
It is useful to analyze the reduction of the BPS conditions in the ABJM model
to those in the super Yang-Mills theory
since much more details of brane configurations are known
in ten-dimensional
 string theory.
The multiple M2-brane effective action is
reduced to that of the D2-branes in type IIA string theory
once one of the transverse direction of the M2-brane world-volume
is compactified on $S^1$.
In the massless ABJM case,
this can be achieved through the novel Higgs mechanism
\cite{Mukhi:2008, Honma:2008jd}.
We briefly summarize this procedure in the following.

First, we assume that the scalar fields $Y^A$ in ABJM model develop a diagonal VEV $v^A \mathbf{1}_N$ and then the
gauge symmetry $U(N) \times U(N)$ is broken down to $U(N)_{\mathrm{diag}}$.
Let us consider fluctuations around the VEV
\renewcommand\arraystretch{1}
\begin{eqnarray}
Y^A = v^A \mathbf{1}_N + \frac{1}{2 |v^A|} (X^{2A+1} + i X^{2A+2}),
\end{eqnarray}
where $X^{2A+1}, X^{2A+2}$ are hermitian matrices in the adjoint
representation of $U(N)_{\mathrm{diag}}$.
Next, we decompose the gauge fields $A_{\mu}, \hat{A}_{\mu}$ as
\begin{eqnarray}
A_{\mu} = a_{\mu} + b_{\mu}, \qquad \hat{A}_{\mu} = a_{\mu} - b_{\mu}.
\end{eqnarray}
The gauge field $a_{\mu}$ becomes dynamical having its kinetic term while $b_{\mu}$ becomes
an auxiliary field.
The algebraic equation of motion for $b_{\mu}$ is solved as
\begin{eqnarray}
b_{\mu} = - \frac{1}{4|v^A|^2}
\left[
\mathcal{D}_{\mu} X^{\phi} - \frac{1}{2} \epsilon_{\mu \nu \rho} f^{\nu \rho}
\right]
+
\mathcal{O} (|v^A|^{-3}),
\label{eq:aux_sol}
\end{eqnarray}
where $\mathcal{D}_{\mu}$ is the covariant derivative with $a_{\mu}$,
$f_{\mu \nu}$ is the field strength of $a_{\mu}$ and
$X^{\phi}$ is defined by
\begin{eqnarray}
X^{\phi} \equiv
\frac{v_A^{\dagger} (X^{2A + 1} + i X^{2A + 2}) - v^A (X^{2A+1} - i
X^{2A+2})}{2 i |v^A|} =
\frac{1}{|\vec v|} (J \vec{v}) \cdot \vec{X},
\end{eqnarray}
where $J = i (\hat \Sigma_{34} + \hat \Sigma_{56} + \hat \Sigma_{78} + \hat \Sigma_{9,10})$ is a complex structure on $\C^4/\Z_k$.
Finally, by taking the limit $v \to \infty,
k \to \infty$ with fixed $1/g^2_{\mathrm{YM}} = \frac{k}{8 \pi |v^A|^2}$,
the procedure substantially leads to the compactification of
the M-theory circle with finite gauge coupling.
Substituting the solution (\ref{eq:aux_sol})
into the action of the ABJM model and taking the limit,
we obtain the action of the $\mathcal{N} = 8$ super Yang-Mills theory
with gauge group $U(N)$ and the gauge coupling $g_{\mathrm{YM}}$.
In the following, we set the VEV in a specific direction $A=4$ for simplicity,
\begin{eqnarray}
v^A = |v| \delta^{A4}, \quad 0 < v \in \mathbb{R}.
\end{eqnarray}
In this case we have $X^{\phi} = X^{10}$. This means the $x^{10}$-direction is the M-theory circle.
The bosonic part of the action becomes
\beq
S_{bosonic} = \frac{1}{g_{\rm YM}^2} \int d^3 x \, \tr \left[ - \frac{1}{2} f_{\mu \nu} f^{\mu \nu} - \D_\mu X^I \D^\mu X^I - \frac{1}{2} [ X^I, X^J ]^2 \right].
\eeq
The supersymmetric variation of the gaugino in three-dimensional
$\mathcal{N} = 8$ super Yang-Mills theory is given by
\begin{eqnarray}
\delta \psi = \partial_{\mu} X_I \Gamma^{\mu} \Gamma^I \xi
+ \frac{1}{2} f_{\mu \nu} \Gamma^{\mu \nu} \Gamma^{10} \xi
+ \frac{i}{2} [X_I, X_J] \Gamma^{IJ} \Gamma^{10} \xi,
\label{D2SUSY}
\end{eqnarray}
where $\Gamma^{\mu} \ (\mu = 0,1,2)$ and $\Gamma^I \ (I = 3, \cdots, 9)$
are ten-dimensional gamma matrices.
The supersymmetry parameter $\xi$ is an $SO(9,1)$ Majorana
spinor that satisfies $\xi = \Gamma^{012} \xi$.

The reduction to ten-dimensions can be applied also to the BPS equations
in the ABJM model.
The reduced conditions are BPS equations in the effective theory of $N$
coincident D2-branes in type IIA string theory.
Let us see this procedure especially focusing on the 1/2 BPS, the most restricted, and 1/12 BPS, the
most generic, equations discussed in sections \ref{sec:1/2BPS}, \ref{sec:BPS_10d} and \ref{sec:BPS}.

\subsection{Reduction of 1/2 BPS equations}
We start from the 1/2 BPS equations discussed in section \ref{sec:1/2BPS}.
There are basically two types of 1/2 BPS equations
corresponding to $\mathcal{A} = \gamma_0 \otimes B$
and $\mathcal{A} = \gamma_2 \otimes C^{(m,n)}$.
The former condition gives the BPS equations
(\ref{eq:BPS0-1}), (\ref{eq:BPS0-2})
supplemented by the Gauss' law (\ref{eq:Gauss1}), (\ref{eq:Gauss2}).
Solutions to these equations are given essentially as point-like object
in the world-volume of M2-branes.
The latter corresponds to the BPS equations
(\ref{eq:BPS1-1}), (\ref{eq:BPS1-2}) with the Gauss' law having fuzzy funnel type solutions.
In the following, we discuss the reduction of
these BPS conditions separately. As we will see, all the 1/2 BPS
equations in the ABJM model reduce to 1/2 BPS equations or the vacuum
condition in D2-brane world-volume theory.

\subsubsection{M2-M2 to D2-F1}
Let us consider M2-branes extending along $x^{9,10}$-directions
which intersect with our fiducial M2-branes at a point.
The projection operator for such configuration is
$\mathcal{A} = \gamma_0 \otimes B$
where $B$ is given by (\ref{B_matrix}).
If we compactify the $x^{10}$-direction in eleven dimensions,
the first M2-branes are reduced to the type IIA fundamental strings
(F1-string) while our fiducial M2-branes become $N$ coincident D2-branes.
Once we apply the reduction procedure to the corresponding BPS equation,
we obtain the following equations,
\beq
f_{12} &=& - i [X^3, X^4] ~=~ - i [X^5, X^6] ~=~ - i [X^7, X^8], \label{eq:D2F1-1} \\
0 &=& \phantom{-} i [X^I, X^J], ~~~~ (I,J) \not= (3,4),\, (5,6),\, (7,8),\, (I,9), \label{eq:D2F1-2}
\eeq
\beq
\D_0 X^I &=& - i [X^9, X^I], ~~~~ \mathcal{D}_1 X^I ~=~ \mathcal{D}_2 X^I ~=~ 0 ~~~ (I=3,\cdots, 8), \label{eq:D2F1-3} \\
\mathcal{D}_0 X^9 &=& 0, ~~~ \mathcal{D}_1 X^9 ~=~ f_{01}, ~~~ \mathcal{D}_2 X^9 ~=~ f_{02}. \label{eq:D2F1-4}
\eeq
On the other hand, the Gauss' law constraint reduces to the following equation
\begin{eqnarray}
\mathcal{D}^{\mu} f_{\mu \nu} - i [X^I, \mathcal{D}_{\nu} X^I] = 0.
\end{eqnarray}
Using the SUSY variation (\ref{D2SUSY}),
we find that the configuration specified by the reduced BPS equations
and the Gauss' law preserve 6 SUSY among 16 SUSY characterized by
\begin{eqnarray}
\Gamma^{09} \Gamma^{10} \xi = - \xi, \qquad
(\Gamma^{12} - \Gamma^{34} - \Gamma^{56} - \Gamma^{78}) \xi = 0.
\end{eqnarray}
Using the conditions (\ref{eq:D2F1-1}), (\ref{eq:D2F1-2}), one can show that all $X^I$ except for $X^9$ commute with each other $[X^I, X^J] = 0$.
Then from (\ref{eq:D2F1-1}), we find that it is possible to choose a gauge $a_1 = a_2 = 0$. The condition (\ref{eq:D2F1-3}) with a gauge $a_0 = - X^9$ implies
\begin{eqnarray}
\p_0 X^I = 0 ~~~\Longrightarrow~~~X^I = \mathrm{diag} (\mathrm{const.}, \cdots, \mathrm{const.}), \quad
(I=3, \cdots, 8).
\end{eqnarray}
In this case, solutions to the equations preserve at least 8 SUSY (1/2 BPS)
determined by
\begin{eqnarray}
\Gamma^{09} \Gamma^{10} \xi = - \xi.
\end{eqnarray}
This is nothing but the projection condition for F1-string extending
along $x^9$-direction \cite{Dabholkar:1990yf}.
Then the solutions are given by
\begin{eqnarray}
a_0 = - X^9, \quad a_1 = a_2 = 0, \qquad (\partial_1^2 + \partial_2^2)
X^9 - [X^I, [X^I, X^9]] = 0.
\label{eq:F1_solution}
\end{eqnarray}
Since $X^I$ are all diagonal, it is easy to find a solution of this equation.
The diagonal harmonic solution of $X^9$ determined by (\ref{eq:F1_solution}) is
known as BIons \cite{Callan:1997kz, Gibbons:1997xz} representing F1-strings ending on the D2-branes.
On the other hand, the off-diagonal parts of $X^9$ have non-trivial solutions, giving more generic configurations of F1,
such as strings stretched between various D2-branes.

It is worthwhile to note that as we have mentioned in section \ref{sec:BPS_10d},
it is possible that KK-monopoles and the M2-branes exist at the same time.
We therefore expect that the reduced BPS equations (\ref{eq:D2F1-1}) - (\ref{eq:D2F1-4})
accommodates D6-branes in addition to the F1-strings. However there seems to be no solutions
corresponding to the D6-branes since all the directions except $X^9$ are constant and do not show the D6-brane behavior.
We discuss this issue in section \ref{Conclusions}.

\subsubsection{M2-M2 to D2-D2}
Next, let us consider the reduction of M2-branes extending along $x^{7,8}$-directions.
Since the directions are different from the one in the M-theory circle, the reduced objects should be D2-branes.
The corresponding projection operator is
\begin{eqnarray}
\Xi = \mathrm{diag} (i \sigma_2, - i \sigma_2, - i \sigma_2).
\end{eqnarray}
It is straightforward to perform the reduction procedure and the result is
the Hitchin equations given by
\beq
f_{12} - i [X^7, X^8] = 0, \qquad (\mathcal{D}_1 + i \mathcal{D}_2)
(X^7 - i X^8) = 0,
\eeq
\begin{eqnarray}
\begin{aligned}
&i [X^I, X^J] = 0, \qquad \mathcal{D}_1 X^I
= \mathcal{D}_2 X^I = 0, ~~ (I \not= 7,8), \\
& f_{01} = f_{02} = 0, \qquad \mathcal{D}_0 X^I = 0, ~~ (I = 3, \cdots, 9).
\end{aligned}
\end{eqnarray}
These solutions preserve 8 SUSY among 16 SUSY, hence the condition
corresponds to 1/2 BPS configuration. The projection condition is
\begin{eqnarray}
\Gamma^{078} \xi = - \xi.
\end{eqnarray}
This is just the condition for D2-branes extending along $x^{0,7,8}$-directions.
Therefore the reduced condition implies D2-branes which intersect
with the fiducial D2-branes.

\subsubsection{$\mathcal{N} = (6,0)$ M2-wave to D2-wave}
Let us consider the $\mathcal{N} = (6,0)$ BPS equations specified by the matrix
\begin{eqnarray}
\Xi = \mathrm{diag} (1,1,1,1,1,1).
\end{eqnarray}
This has been discussed in section \ref{sec:BPS_10d}. The corresponding
object is an M-wave extending along $x^{0,1}$-directions shared with the fiducial M2-branes.
After the reduction, the object becomes the type IIA wave in ten dimensions.
The BPS equation reduces to
\beq
f_{02} - f_{12} = 0, \qquad (\mathcal{D}_0 - \mathcal{D}_1) X^I = 0,
\eeq
\begin{eqnarray}
\begin{aligned}
& \mathcal{D}_2 X^9 = i [X^3, X^4] = i [X^5, X^6] = i [X^7, X^8], \\
& [X^I, X^J] = 0, ~~~ (I,J) \not= (3,4), (5,6), (7,8), \\
& f_{01} = 0, \hs{5} \mathcal{D}_2 X^I = 0, ~~~ (I=3, \cdots, 8).
\end{aligned}
\end{eqnarray}
As we have noticed, it is possible to show that there is no non-trivial
(non-diagonal constant) solution that satisfies $[X^I, X^J] \not=0$.
Therefore preserved supersymmetry is actually 8 SUSY determined by
\begin{eqnarray}
\Gamma^{01} \xi = \xi,
\end{eqnarray}
which is $\mathcal{N}=(8,0)$ SUSY in $(1 \! + \! 1)$-dimensional theory
\footnote{This $(1+1)$ dimensional directions are common world-volume shared by both the waves (or other BPS objects) and the fiducial D2-branes.
The statement "$\mathcal{N} = (m,n)$" should be understood as the number of (anti)chiral supercharges that are preserved by the effective theory living in the common world-volume.
}.

\subsubsection{$\mathcal{N} = (5,1)$ condition to vacuum}
Next, we consider the $\mathcal{N} = (5,1)$ BPS equations.
In general, the corresponding objects in M-theory is M5-branes extending along the calibrated submanifold in $\mathbb{C}^4/\mathbb{Z}_k$
specified by the 4-form (\ref{eq:51_form}).
We can show that these BPS equations reduce to vacuum conditions in super Yang-Mills theory.
This would mean that the corresponding objects in ten dimensions do not exist or it is non-BPS configuration.

\subsubsection{$\mathcal{N} = (4,2)$ M2-M5 to D2-D4}
Let us consider $\mathcal{N} = (4,2)$ BPS condition representing
M5-branes that extend along $x^{0,2,5,6,7,10}$-directions.
The corresponding projection operator is
\begin{eqnarray}
\Xi = (1,1,1,1,-1,-1).
\end{eqnarray}
Once we compactify $x^{10}$-directions, it correspond to direct
dimensional reduction of the M5-branes leaving D4-branes in ten
dimensional space-time. The reduced BPS equations are the Nahm equations
\beq
\mathcal{D}_2 X^5 = i [X^6, X^9], \quad \mathcal{D}_2 X^6 = i [X^9,
X^5], \quad \mathcal{D}_2 X^9 = i[X^5, X^6],
\eeq
\begin{eqnarray}
\begin{aligned}
& f_{01} = f_{02} = f_{12} = \mathcal{D}_0 X^I = \mathcal{D}_1 X^I = 0, \\
& \mathcal{D}_2 X^I = i [X^I, X^J] = 0, ~~~ (I =3,4,7,8,~ J = 3, \cdots, 9).
\end{aligned}
\end{eqnarray}
The solutions preserve 8 SUSY determined by
\begin{eqnarray}
\Gamma^{2569} \Gamma^{10} \xi = \xi \quad \leftrightarrow \quad \Gamma^{01569}
\Gamma^{10} \xi = \xi.
\end{eqnarray}
This is nothing but the projection condition of D4-branes extending
along $x^{0,1,5,6,9}$-directions.
The preserved SUSY is $\mathcal{N}=(4,4)$ in $(1 \! + \! 1)$-dimensions.

\subsubsection{$\mathcal{N} = (3,3)$ M2-M5 to D2-D4}
Consider $\mathcal{N} = (3,3)$ BPS condition
that corresponds to M5-branes extending along $x^{0, 2,5,6,9, 10}$-directions
for which the projection operator is given by
\begin{eqnarray}
\Xi = \mathrm{diag} (-1,1,1,-1,-1,1).
\end{eqnarray}
The BPS equations are reduced to the following Nahm equations
\beq
\mathcal{D}_2 X^4 = i [X^6,X^8], \qquad \mathcal{D}_2 X^6 = i [X^8, X^4],
\qquad \mathcal{D}_2 X^8
 = i [X^4, X^6],
\eeq
\begin{eqnarray}
\begin{aligned}
& f_{01} = f_{02} = f_{12} = \mathcal{D}_0 X^I = \mathcal{D}_1 X^I = 0, \\
& \mathcal{D}_2 X^I = i [X^I, X^J], ~~~ (I = 3,5,7,9, ~J = 3, \cdots, 9).
\end{aligned}
\end{eqnarray}
The solution should represent D4-branes in ten dimensions. Actually, the
preserved supersymmetry is characterized by
\begin{eqnarray}
\Gamma^{2468} \Gamma^{10} \xi = \xi \quad \leftrightarrow \quad \Gamma^{01468} \Gamma^{10} \xi = \xi,
\end{eqnarray}
and represents D4-branes in $x^{0,1,4,6,8}$-directions.
The projection condition keeps $\mathcal{N}=(4,4)$ SUSY in $(1 \! + \! 1)$-dimensions.

\subsection{Reduction of 1/12 BPS equations}
So far, we have studied the reduction of the 1/2 BPS configurations in
the ABJM model.
There are several ways to find BPS equations that keep less than 6 SUSY.
We have shown in section \ref{sec:BPS}
that if some of intersecting branes in the 1/2 BPS configurations have
non-trivial intersecting angles with orbifolded planes, the preserved 6 SUSY is further broken. In
addition to this mechanism, once one imposes several kind of projection
conditions or restricting the form of the projection matrices $\Xi$, the remaining supersymmetries can be reduced down to 1 SUSY among
12 SUSY. In the following subsections, we consider 1/12 BPS conditions derived in this way
and see what kind of structures are obtained after the
dimensional reduction to ten dimensions. We will see that the 1/12 BPS
equations in the ABJM model reduce to 1/16 BPS equations in $\mathcal{N}
= 8$ super Yang-Mills theory.

\subsubsection{Intersecting M5-branes}
Let us consider $\mathcal{N} = (1,0)$ BPS equations
with the projection operator
\begin{eqnarray}
\Xi = \mathrm{diag} (1,0,0,0,0,0).
\end{eqnarray}
This corresponds to the fiducial M2-branes ending on several (anti) M5-branes together with M-waves.
Performing the reduction, we obtain the following reduced equations
\begin{eqnarray}
\begin{aligned}
& f_{02} = f_{12}, ~~ f_{01} = 0 \qquad (\mathcal{D}_0 - \mathcal{D}_1) X^I = 0, ~~~~~\mathcal{D}_2 X^I = i g_{IJK} [X^J, X^K],
\end{aligned}
\end{eqnarray}
where the anti-symmetric structure constants $g_{IJK}$ are defined by
\begin{eqnarray}
g_{349} = g_{358} = - g_{367} = - g_{457} = - g_{468} =  g_{569} = -
g_{789} = 1/2.
\end{eqnarray}
This configuration preserves 1 SUSY, hence it is 1/16 BPS condition determined
by
\begin{eqnarray}
\Gamma^{2358} \Gamma^{10} \xi = \xi, ~~~
\Gamma^{2457} \Gamma^{10} \xi = - \xi, ~~~
\Gamma^{2367} \Gamma^{10} \xi = - \xi, ~~~
\Gamma^{01} \xi = \xi.
\end{eqnarray}
These conditions imply co-existence of D4-branes with type IIA-waves.

\subsubsection{Intersecting M2-M5-branes}
Let us consider BPS equations specified by
\renewcommand\arraystretch{1.5}
\beq
\Xi = \ba{c|c} \sigma_3 & \\ \hline  & \hspace{1mm} \mbox{{\Huge $\ast$}}_4 \ea, \hs{10}
\Xi' = \ba{c|c} \sigma_1 & \\ \hline  & \hspace{1mm} \mbox{{\Huge $\ast$}}_4 \ea.
\eeq
The reduced equations are given by
\renewcommand\arraystretch{1}
\beq
f_{01} = \mathcal{D}_1 X^9, \quad f_{02} = \mathcal{D}_2 X^9, \quad f_{12} = - [Z^{\hat A}, Z_{\hat A}^\dagger] - [ W, W^\dagger],
\eeq
\beq
\mathcal{D}_0 Z^{\hat A} = i [Z^{\hat A}, X^9], \hs{10}
\mathcal{D}_0 W = i [W, X^9],
\eeq
\beq
\mathcal{D}_{\bar{z}} Z^{\hat{A}} = \frac{i}{2} \epsilon^{\hat{A}
\hat{B}} [Z^{\dagger}_{\hat{B}}, W^\dagger], \hs{10}
\mathcal{D}_{\bar z} W = \frac{i}{4} \epsilon^{\hat{A} \hat{B}}
[Z_{\hat{A}}^\dagger, Z_{\hat{B}}^\dagger],
\eeq
where we have defined $z = x^1 + i x^2$, $\mathcal{D}_z = \frac{1}{2}
(\mathcal{D}_1 - i \mathcal{D}_2)$ and
\begin{eqnarray}
Z^1 = X^3 + i X^4, \hs{5} Z^2 = X^5 + i X^6, \hs{5} W = X^7 - i X^8.
\end{eqnarray}
This configuration preserves 1 SUSY. Therefore it is 1/16 BPS condition
specified by
\begin{eqnarray}
\Gamma^{1357} \Gamma^{10} \xi = \xi, \quad
\Gamma^{2358} \Gamma^{10} \xi = \xi, \quad
\Gamma^{1458} \Gamma^{10} \xi = \xi, \quad
\Gamma^{1368} \Gamma^{10} \xi = \xi, \quad
\end{eqnarray}
These conditions imply the existence of the multiple kinds of D2 and D4-branes.

\section{Conclusions and discussions} \label{Conclusions}

In this paper, we have classified
BPS equations in the ABJM model
derived from the vanishing conditions of $\mathcal{N} = 6$ supersymmetry transformation of the fermions.
The BPS equations are characterized
in terms of the unbroken supercharges
specified by the projection conditions
for the spinor parameters of the supersymmetry transformation.
For the 1/2 BPS case, we have found
the two types of the projection conditions
\eqref{half_BPS1} and \eqref{half_BPS2},
which correspond to co-dimension two and one objects, respectively.

The analysis of the projection conditions for the supersymmetry parameter
provides not only classifications of BPS equations in the ABJM model,
but also an insight into the BPS objects
extending in eleven-dimensional space-time.
We have discussed the projection conditions
for eleven-dimensional spinor parameters that are
consistent with the existence of our fiducial M2-branes in $\C^4/\Z_k$ orbifold.
Those projection conditions give the information on
how BPS objects extend in eleven-dimensional space-time.
We have made the mapping \eqref{operator_mapping}
between the projection matrices
in eleven dimensions and those in the ABJM model,  
and found the eleven-dimensional interpretation of
the BPS objects in the ABJM model.

Starting from the 1/2 BPS conditions, we have also obtained $n/12 \ (n=1, \cdots, 5)$ BPS conditions.
A careful investigation of the SUSY breaking pattern
allows us to establish the correspondence between each BPS equations
and the possible brane configurations in M-theory.
Those include M2, M5-branes which have non-trivial angles with the orbifolded planes and several bunches of M2, M5-branes filling subspaces in the internal space $\mathbb{C}^4/\mathbb{Z}_k$.
We have also shown that the existence of various M-theoretical objects, such as M-waves, KK-monopoles and M9-branes, is consistent with the BPS conditions.
These results are summarized in Table \ref{tab:summary}.

To see the dimensional reduction of the BPS equations
is another consistency check.
By taking the reduction limit via the novel Higgs mechanism
the BPS equations in the ABJM model reduce to
those in the three-dimensional $\mathcal{N} = 8$ super Yang-Mills theory
in a consistent way with the reduction from M-theory to Type IIA string theory.
Our results reveal strong evidences that the ABJM model correctly captures dynamics of multiple M2-branes.

It is important to bear in mind that we discussed here the relation
between the BPS equations in the ABJM model and
the BPS objects in M-theory only
in the aspects of the SUSY conditions.
Of course the classification of
the projection condition for SUSY parameters
is not sufficient to conclude the correspondence
between the BPS equations and the M-theoretical objects
since the derived equations may have no non-trivial solutions.
This is a conceivable argument since, as we have claimed in the previous section,
there seem to be no non-trivial solutions in ten-dimensional BPS equations that
correspond to D6, D8 and NS5-branes.
This fact indicates that some of BPS equations in ABJM model have only trivial solutions.
To complete the analysis, we should confirm the existence of
solutions of the BPS equations, and examine
whether the solutions have appropriate property as
M2-brane, M5-brane, and so on.
We have found a number of non-trivial solutions for the BPS equations derived
in this paper. The detail discussions and explicit forms of these solutions
will be found in the forthcoming paper \cite{inpreparation}.

\begin{table}[ht]
\begin{center}
\begin{tabular}{|c|c|c|}
\hline
$\mathcal{N}$ & Residual symmetry & Intersecting branes \\
\hline
6 & $SU (3) \times U (1)^2$ & M2, KK-monopoles \\
6 & $SU (4)$ & M2, M9, M-waves \\
6 & $SU (2) \times SU (2) \times \rm{U}(1)^2 $ & M5 \\
5 & $SU (2) \times SU (2)$ & M5 with angles\\
4 & $U (1)^3 $ & M5 with angles \\
4 & $SU (2) \times U (1)^2$ & M2, KK-monopoles \\
4 & $SU (2) \times U (1)^2$ & M2 with angles  \\
3 & $SO (3)$ & M2 ending on M5 \\
2 & $SU (2) \times SU (2) \times U (1)^2 $ & M2, KK-monopoles \\
2 & $U (1) \times U (1)$ & M2 ending on M5 \\
1 & $SU (2) \times SU(2)$ & M2 ending on M5, M9, M-waves\\
$ n+m $ & $ Spin (n) \times Spin (m) \times Spin (6-n-m)$ & M5, M9, M-waves \\
\hline
\end{tabular}
\caption{Classification of the BPS equations in
the number of preserved supercharges, the symmetry of BPS equations and
the corresponding M-theoretical objects. $\mathcal{N}$ is the number of preserved supercharges.}
\label{tab:summary}
\end{center}
\end{table}

\subsection*{Acknowledgments}
K.~I. gratefully acknowledges the financial support from the Global Center of Excellence Program by the Ministry of Education, Culture, Sports, Science and Technology (MEXT) of Japan through the ``Nanoscience and Quantum Physics" Project of the Tokyo Institute of Technology, and support from the Iwanami Fujukai Foundation. This work is also supported by the Grant-in-Aid for the Global Center of Excellence Program of the Kyoto University ``The Next Generation of Physics, Spun from Universality and Emergence" from MEXT.
The work of K.~I and S.~S is supported by the Japan Society for the Promotion of Science (JSPS) Research Fellowship.
\begin{appendix}

\section{BPS equations}\label{appendix:BPSeqs}
In this appendix, we summarize the BPS equations discussed in this paper.
In order to write down the BPS equations,
it is convenient to define $\beta_A^{BC}$ by
\beq
\beta_A^{BC} &\equiv& Y^B Y_A^\dagger Y^C - Y^C Y_A^\dagger Y^B.
\eeq
As we have mentioned in section \ref{sec:1/2BPS}, all the BPS conditions in
this Appendix are supplemented by the Gauss' law constraints (\ref{eq:Gauss1}),
(\ref{eq:Gauss2}). This is the necessary condition that solutions of the
BPS equations satisfy the equations of motion.

\subsection{M2-branes}
Let us consider the BPS configurations of intersecting M2-branes preserving
the supercharges determined by the following condition
\beq
\gamma_0 \Xi_{ij} \epsilon_j = \epsilon_i,
\eeq
where $\Xi$ is a 6-by-6 real anti-symmetric matrix
satisfying $\Xi^2 = -\mathbf 1_6$.
The BPS equations given below preserve $2n~(n=1,2,3)$ supercharges
and specified by the following matrices
\beq
\Xi = \left\{
\begin{array}{ll}
\diag ( i \sigma_2 , \,\ast_2\, , \,\ast_2\, ) & \mbox{for}~n=1 \\
\diag ( i \sigma_2 , i \sigma_2 , \,\ast_2\, ) & \mbox{for}~n=2 \\
\diag ( i \sigma_2 , i \sigma_2 , i \sigma_2 ) & \mbox{for}~n=3 \\
\end{array} \right..
\eeq
The corresponding BPS equations take the following form
\beq
- D_0 Y^B (\Gamma_j)_{BA} \Xi_{ji} + \Upsilon_A^{BC} (\Gamma_i)_{BC} = 0, \hs{10} D_1 Y^B (\Gamma_i)_{BA} + D_2 Y^B (\Gamma_j)_{BA} \Xi_{ji} = 0,
\eeq
where $1 \leq i,j \leq 2n$. The BPS configurations saturate the BPS bound for the energy
\beq
E \geq \frac{k}{4\pi n} \int d^2 x \, \epsilon^{MN} \p_M \tr \left[ Y_C^\dagger D_N Y^B \right] (\Gamma_i \Gamma_j^\dagger \Xi_{ji})_B{}^C, \hs{10} (M,N=1,2),
\eeq
and if the BPS equations are satisfied, the energy becomes
\beq
E = \frac{k}{2\pi} \int d^2 x \, \p_M \tr \left[ Y_A^\dagger D^M Y^A \right].
\eeq
\vs{3}
In the following, we present detail structures of each $n=1,2,3$ BPS
conditions.

\begin{itemize}
\item 1/6 BPS equations \\
For the 1/6 BPS equations of intersecting M2-branes,
the anti-symmetric matrix $\Xi$ is given by
\renewcommand\arraystretch{1.5}
\beq
\Xi = - \ba{c|c}
i \sigma_2 & \\ \hline
& \phantom{(} \mbox{{\Huge $\ast$}}_4 \ea.
\eeq
Since the form of the matrix $\Xi$ is invariant under $SO(4) \times SO(2)$,
the corresponding BPS equations have $SU(2) \times SU(2) \times U(1)$ symmetry
for which $Y^a~(a=1,2)$ and $Y^{\dot a}~(\dot a = 3,4)$ are in $(\mathbf 2,\mathbf 1)_{1}$ and $(\mathbf 1,\mathbf 2)_{-1}$ respectively. The BPS equations are given by
\renewcommand\arraystretch{1}
\beq
D_0 Y^a = i (\beta_b^{a b} - \beta_{\dot b}^{a \dot b}), \hs{5}
(D_1 - i D_2) Y^a = 0, \\
D_0 Y^{\dot a} = i ( \beta_b^{\dot a b} - \beta_{\dot b}^{\dot a \dot b}), \hs{5}
(D_1 + i D_2) Y^{\dot a} = 0,\,
\label{eq:M2_1/6_a}
\eeq
\vs{-7}
\beq
\beta_a^{\dot a \dot b} ~=~ \beta_{\dot a}^{ab} ~=~ 0.
\label{eq:M2_1/6_b}
\eeq

\item 1/3 BPS equations \\
For the 1/3 BPS equations of intersecting M2-branes,
the anti-symmetric matrix $\Xi$ is given by
\beq
\Xi = \ba{c|c|c}
i \sigma_2 & & \\ \hline
& i \sigma_2 & \\ \hline
& & \mbox{{\Large $\ast$}}_2 \ea.
\eeq
The corresponding BPS equations are symmetric under $SU(2) \times U(1)^2$
for which $Y^\alpha~(\alpha=1,3)$, $Y^2$, $Y^4$ are in $\mathbf 2_{(1,1)}$, $\mathbf 1_{(-2,0)}$, $\mathbf 1_{(0,-2)}$, respectively. The BPS equations are given by
\beq
D_0 Y^\alpha = i (\beta_2^{\alpha 2} - \beta_4^{\alpha 4}), \hs{5} D_1 Y^\alpha = D_2 Y^\alpha = 0,
\eeq
\vs{-7}
\beq
D_0 Y^2 = - i \beta_4^{24}, \hs{5} (D_1 - i D_2) Y^2 = 0, \\
D_0 Y^4 = \phantom{-} i \beta_2^{42}, \hs{5} (D_1 + i D_2) Y^4 = 0,\,
\label{eq:M2_1/3_a}
\eeq
\vs{-7}
\beq
\beta_\alpha^{\beta \gamma}=0, \hs{7}
\beta_2^{4 \alpha} = \beta_4^{2 \alpha} = 0, \hs{10}
\beta_\alpha^{2\beta} = \frac{1}{2} \delta_\alpha^\beta \beta_\gamma^{2 \gamma}, \hs{5}
\beta_\alpha^{4\beta} = \frac{1}{2} \delta_\alpha^\beta \beta_\gamma^{4\gamma}.
\label{eq:M2_1/3_b}
\eeq

\item 1/2 BPS equations \\
For the 1/2 BPS equations of intersecting M2-branes,
the anti-symmetric matrix $\Xi$ is given by
\beq
\Xi = - \ba{c|c|c}
i \sigma_2 & & \\ \hline
& i \sigma_2 & \\ \hline
& & i \sigma_2 \ea.
\eeq
The corresponding BPS equations have $SU(3) \times U(1)^2$
for which $Y^i~(i=1,2,3)$ and $Y^4$ are in $\mathbf 3_{1}$ and $\mathbf 1_{-3}$, respectively. The BPS equations are given by
\beq
D_0 Y^i = -i \beta_4^{i4}, \,\hs{5} D_1 Y^i = D_2 Y^i = 0,
\label{eq:M2_1/2_a}
\\
D_0 Y^4 = \phantom{-} \frac{i}{3} \beta_i^{4i}, \hs{5} (D_1 + i D_2) Y^4 = 0,
\eeq
\vs{-7}
\beq
\beta_i^{jk} ~=~ \beta_4^{jk} ~=~ 0, \hs{10} \beta_i^{4j} = \frac{1}{3} \delta_i^j \beta_k^{4k}.
\label{eq:M2_1/2_b}
\eeq

\end{itemize}

\subsection{M5-branes}
We summarize here the complete list of BPS equations
for M5-branes in order of increasing unbroken supercharges.
The unbroken supercharges are specified by
\beq
\gamma_2 \Xi_{ij} \epsilon_j = \epsilon_i,
\eeq
where $\Xi$ is a 6-by-6 real symmetric matrix
satisfying $\Xi^2 = \mathbf 1_6$.
For the BPS configurations with $\mathcal N = (m,n)$ supersymmetry,
the symmetric matrix $\Xi$ takes the form
\renewcommand\arraystretch{1.5}
\beq
\Xi = \ba{c|c|c}
\mathbf 1_m & & \phantom{\Big(} \\ \hline
& -\mathbf 1_n & \phantom{\Big(} \\ \hline
& & \, \mbox{{\Huge $\ast$}}_{6-m-n} \ea.
\eeq
Since the form of the matrix is invariant under
$SO(m) \times SO(n) \times SO(6-m-n)$ subgroup of $SO(6)_R$,
the corresponding BPS equations have
$Spin(m) \times Spin(n) \times Spin(6-m-n) \subset SU(4)_R$ symmetry.
The $\mathcal{N}=(m,n)$ BPS equations
are given by the following general formula;
\renewcommand\arraystretch{1}
\beq
D_0 Y^A = \pm D_1 Y^A,~~~(\mbox{if $n=0$ or $m=0$}), \hs{7} D_0 Y^A = D_1 Y^A = 0,~~~(\mbox{if $m, n \not = 0$}),
\eeq
\beq
D_2 Y^A &=& - \Upsilon ^{BC}_{D}(\Gamma_{i})_{BC}(\Gamma^{\dagger}_{i})^{DA}, \hs{7} (\mbox{no sum over $i$,}~~~i=1,\cdots,m), \\
D_2 Y^{A} &=& \phantom{-} \Upsilon ^{BC}_{D}(\Gamma_{j})_{BC}(\Gamma^{\dagger}_{j})^{DA}, \hs{7} (\mbox{no sum over $j$,}~~~j=m+1,\cdots,m+n).
\eeq
The BPS configurations saturate the following BPS bound for the energy
\beq
E \geq \pm P_1 + \frac{k}{4\pi(m+n)} \int d^2 x \, \p_2 \tr \left[ Y_A^\dagger \Upsilon_D^{BC} \right] \sum_{i,j=1}^{m+n} (\Gamma_i)_{BC} (\Gamma_j^\dagger)^{AD} \Xi_{ij} ,
\eeq
where $P_1$ is the conserved momentum given by
\beq
P_1 \equiv \frac{k}{2\pi} \int d^2 x \, \tr \left[ D_0 Y_A^\dagger D_1 Y^A + D_1 Y_A^\dagger D_0 Y^A \right].
\eeq
For any configurations satisfying the BPS equations, the energy is given by
\beq
E = \pm P_1 + \frac{k}{4\pi} \int d^2 x \, \p_2 \tr \left[ Y_A^\dagger D_2 Y^A \right].
\eeq

\paragraph{$1/12$ BPS equations}
\begin{itemize}
\item $\mathcal{N}=(1,0)$ \\
The matrix $\Xi$ for the 1/12 BPS configurations
is invariant under $SO(5)$,
so that the corresponding BPS equations have
$Spin(5) \cong USp(4)$ symmetry defined by
\beq
U^T \Gamma_1 U = \Gamma_1, \hs{10} U U^\dagger = \mathbf 1_4.
\eeq
The scalar fields $Y^A~(A=1,2,3,4)$ are in $\mathbf 4$ of $USp(4)$.
The BPS equations are given by
\beq
D_2 Y^A = \beta_D^{AD} + (\Gamma_1^\dagger)^{AB} (\Gamma_1)_{CD} \beta_B^{CD} , \hs{10} (D_0-D_1) Y^A = 0.
\eeq
\end{itemize}
\paragraph{$1/6$ BPS equations}
\begin{itemize}
\item $\mathcal{N}=(2,0)$ \\
For the BPS configurations with $\mathcal N = (2,0)$ preserved supersymmetry,
the symmetry of the corresponding BPS equations is
$SU(2) \times U(1) $ for which
$Y^a~(a=1,2)$ and $Y^{\dot a}~(\dot a = 3,4)$ are
in $(\mathbf 2, \mathbf 1)_1$ and $(\mathbf 1, \mathbf 2)_{-1}$, respectively.
The BPS equations are given by
\beq
D_2 Y^{a} = - \beta^{ab}_{b} + \beta^{a \dot{b}}_{\dot{b}}, \hs{10} (D_0-D_1) Y^a =0, \\
D_2 Y^{\dot{a}} = - \beta^{\dot{a} \dot{b}}_{\dot{b}} + \beta^{\dot{a} b}_{b}, \hs{10} (D_0-D_1) Y^{\dot a} =0.\,
\eeq
\vs{-7}
\beq
\beta^{\dot{c} \dot{d}}_{b} = \beta^{c d}_{\dot{b}} = 0.
\eeq

\item $\mathcal{N}=(1,1)$ \\
For the BPS configurations with $\mathcal N = (1,1)$ preserved supersymmetry,
the symmetry of the corresponding BPS equations is
$SU(2) \times SU(2)$ for which
$Y^a~(a=1,2)$ and $Y^{\dot a}~(\dot a = 3,4)$ are
in $(\mathbf 2, \mathbf 1)$ and $(\mathbf 1, \mathbf 2)$, respectively.
The BPS equations are given by
\beq
D_2 Y^a = - \epsilon^{ab} \epsilon_{\dot{c} \dot{d}} \beta^{\dot{c} \dot{d}}_{b}, \hs{10} D_0 Y^a = D_1 Y^a = 0, \\
D_2 Y^{\dot{a}} = - \epsilon^{\dot{a} \dot{b}} \epsilon_{c d} \beta^{c d}_{\dot{b}}, \hs{10} D_0 Y^{\dot a} = D_1 Y^{\dot a} = 0.\,
\eeq
\vs{-7}
\beq
\beta^{ab}_{b} = \beta^{a \dot{b}}_{\dot{b}}, \hs{10}
\beta_{\dot{b}}^{\dot{a} \dot{b}} = \beta_{b}^{\dot{a} b}.
\eeq
\end{itemize}

\paragraph{$1/4$ BPS equations}
\begin{itemize}
\item $\mathcal{N}=(3,0)$ \\
For the BPS configurations with $\mathcal N = (3,0)$ preserved supersymmetry,
the symmetry of the corresponding BPS equations is $SO(4)$ defined by
\beq
U^T g \, U = g, \hs{10} U \in SU(4)_R,
\eeq
where the $SO(4)$ invariant tensor $g$ is given by
\beq
g_{AB} \equiv \ba{c|c} & i \sigma_2 \\ \hline -i \sigma_2 & \ea.
\eeq
The scalar fields $Y^A~(A=1,2,3,4)$ are in $\mathbf 4$ of $SO(4)$.
The BPS equations are given by
\beq
D_2 Y^A = - \frac{1}{3} ( \epsilon^{ABCD} \beta_{BCD} - \beta_D^{AD} ), \hs{10} (D_0 - D_1) Y^A = 0,
\eeq
\beq
\beta_{ABC} = \beta_{[ABC]} + \frac{1}{2} \epsilon_{DE[AB} \beta_{C]}^{DE} - \frac{2}{3} g_{A[B} g_{C]E} \beta_D^{ED} - \frac{1}{3} \epsilon_{BC}{}^{DE} (\beta_{DAE} + \beta_{ADE}).
\eeq
Here, we have used the $SO(4)$ invariant tensor $g$ to lower the indices
\beq
\beta_{ABC} \equiv g_{BB'} g_{CC'} \beta_A^{B'C'}, \hs{10}
\epsilon_{BC}{}^{DE} \equiv g_{BB'} g_{CC'} \epsilon^{B'C'DE}.
\eeq

\item $\mathcal{N}=(2,1)$ \\
For the BPS configurations with $\mathcal N = (2,1)$ preserved supersymmetry,
the symmetry of the corresponding BPS equations is
$SU(2) \times U(1)$ for which
$Y^a~(a=1,2)$ and $Y^{\dot a}~(\dot a = 3,4)$ are
in $\mathbf 2_1$ and $\mathbf 2_{-1}$, respectively.
The BPS equations are given by
\beq
D_{2} Y^{a} = - \beta^{ab}_{b} + \beta^{a \dot{b}}_{\dot{b}}, \hs{10} D_0 Y^a = D_1 Y^a =0, \\
D_{2} Y^{\dot{a}} = - \beta^{\dot{a} \dot{b}}_{\dot{b}} + \beta^{\dot{a} b}_{b}, \hs{10} D_0 Y^{\dot a} = D_1 Y^{\dot a} =0.\,
\eeq
\beq
\beta^{\dot{c} \dot{d}}_{b} = \beta^{c d}_{\dot{b}} = 0, \hs{10} \sum_{b=1}^2 \beta_b^{a, b+2} = \sum_{\dot b=3}^4 \beta_{\dot b}^{a, \dot b-2} = 0.
\eeq

\end{itemize}

\paragraph{$1/3$ BPS equations}
\begin{itemize}
\item $\mathcal{N}=(4,0)$ \\
For the BPS configurations with $\mathcal N = (4,0)$ preserved supersymmetry,
the symmetry of the corresponding BPS equations is
$SU(2) \times SU(2) \times U(1)$ for which
$Y^\alpha~(a=1,3)$ and $Y^{\dot \alpha}~(\dot \alpha = 2,4)$ are
in $(\mathbf 2, \mathbf 1)_1$ and $(\mathbf 1, \mathbf 2)_{-1}$, respectively.
The BPS equations are given by
\beq
D_2 Y^\alpha = \beta^{\alpha \beta}_{\beta}, \hs{10} (D_0 - D_1) Y^\alpha = 0, \\
D_2 Y^{\dot \alpha} = \beta^{\dot \alpha \dot \beta}_{\dot \beta}, \hs{10} (D_0 - D_1) Y^{\dot \alpha} = 0.\,
\eeq
\beq
\beta_\alpha^{\dot \alpha \beta} = \frac{1}{2} \delta_\alpha^\beta \beta_\gamma^{\dot \alpha \gamma}, \hs{10} \beta_{\dot \alpha}^{\alpha \dot \beta} = \frac{1}{2} \delta_{\dot \alpha}^{\dot \beta} \beta_{\dot \gamma}^{\alpha \dot \gamma}.
\eeq

\item $\mathcal{N}=(3,1)$ \\
For the BPS configurations with $\mathcal N = (3,1)$ preserved supersymmetry,
the symmetry of the corresponding BPS equations is
$SU(2) \times U(1)$ for which
$Y^\alpha~(\alpha=1,3)$ and $Y^{\dot \alpha}~(\dot \alpha = 2,4)$ are
in $\mathbf 2_1$ and $\mathbf 2_{-1}$, respectively.
The BPS equations are given by
\beq
D_2 Y^\alpha = 2 \beta^{\alpha \beta}_{\beta}, \hs{10} D_0 Y^\alpha = D_1 Y^\alpha = 0, \\
D_2 Y^{\dot \alpha} = 2 \beta^{\dot \alpha \dot \beta}_{\dot \beta}, \hs{10} D_0 Y^{\dot \alpha} = D_1 Y^{\dot \alpha} = 0.\,
\eeq
\beq
\beta_\alpha^{\dot \beta \beta} = \frac{1}{2} \delta_\alpha^\beta (\beta_\gamma^{\dot \beta \gamma} + \beta_{\dot \gamma}^{\dot \beta \dot \gamma} ) - \delta_\alpha^{\dot \beta-1} \beta_{\dot \gamma}^{\beta+1, \dot \gamma}, \\
\beta_{\dot \alpha}^{\beta \dot \beta} = \frac{1}{2} \delta_{\dot \alpha}^{\dot \beta} (\beta_{\dot \gamma}^{\beta \dot \gamma} + \beta_{\gamma}^{\beta \gamma} ) - \delta_{\dot \alpha}^{\beta+1} \beta_{\gamma}^{\dot \beta-1, \gamma}.\,
\eeq

\item $\mathcal{N}=(2,2)$ \\
For the BPS configurations with $\mathcal N = (2,2)$ preserved supersymmetry,
the symmetry of the BPS equations is
$U(1)^3$ corresponding to the Cartan subalgebra of $SU(4)_R$.
The BPS equations are given by
\beq
D_2 Y^a = - ( \beta_b^{a b} - \beta_{\dot b}^{a \dot b} ), \hs{10} D_0 Y^a = D_1 Y^a = 0, \\
D_2 Y^{\dot a} = - ( \beta_{\dot b}^{\dot a \dot b} - \beta_b^{\dot a b} ), \hs{10} D_0 Y^{\dot a} = D_1 Y^{\dot a} = 0,\,
\eeq
\begin{eqnarray}
\beta^{\dot a \dot b}_{b} = \beta^{a b}_{\dot{b}} = 0, \hs{10}
\beta^{a 3}_{1} = \beta^{a 4}_{2}  = \beta^{\dot{a} 1}_{3} = \beta^{\dot a 2}_{4} = 0,
\end{eqnarray}
where the indices run as $a=1,2, \dot{a} = 3,4$.
\end{itemize}

\paragraph{$5/12$ BPS equations}
\begin{itemize}
\item $\mathcal{N}=(5,0)$ \\
The matrix $\Xi$ for the 1/12 BPS configurations
is invariant under $SO(5)$,
so that the corresponding BPS equations have
$Spin(5) \cong USp(4)$ symmetry defined by
\beq
U^T \Gamma_6 U = \Gamma_6, \hs{10} U \in SU(4).
\eeq
The scalar fields $Y^A~(A=1,2,3,4)$ are in $\mathbf 4$ of $USp(4)$.
The BPS equations are given by
\beq
D_2 Y^A = \frac{1}{5} \left( \beta_D^{AD} - (\Gamma_6^\dagger)^{AD} (\Gamma_6)_{BC} \beta_D^{BC} \right), \hs{10} (D_0 - D_1) Y^A = 0.
\eeq
\beq
\beta_A^{BC} = - \frac{1}{4} \left[ \delta_A^D (\Gamma_6^\dagger)^{BC} (\Gamma_6)_{E F} + 2 (\Gamma_6)_{AF} (\Gamma_6^\dagger)^{D[B} \delta_{E}^{C]} + 2 \delta_A^{[B} \delta_{E}^{C]} \delta_{F}^{D} \right] \beta_D^{E F}.
\eeq

\item $\mathcal{N}=(4,1)$ \\
For the BPS configurations with $\mathcal N = (4,1)$ preserved supersymmetry,
the symmetry of the corresponding BPS equations is
$SU(2) \times SU(2)$ for which
$Y^{\alpha}~(\alpha =1,3)$ and $Y^{\dot \alpha}~(\dot \alpha = 2,4)$ are
in $(\mathbf 2, \mathbf 1)$ and $(\mathbf 1, \mathbf 2)$, respectively.
The BPS equations are given by
\beq
D_2 Y^\alpha = \beta^{\alpha \beta}_{\beta}, \hs{10} D_0 Y^\alpha = D_1 Y^\alpha = 0, \\
D_2 Y^{\dot \alpha} = \beta^{\dot \alpha \dot \beta}_{\dot \beta}, \hs{10} D_0 Y^{\dot \alpha} = D_1 Y^{\dot \alpha} = 0.\,
\eeq
\beq
\beta_\alpha^{\dot \alpha \dot \beta} = \epsilon_{\alpha \beta} \epsilon^{\dot \gamma [\dot \alpha} \beta_{\dot \gamma}^{\dot \beta] \beta}, \hs{10}
\beta_{\dot \alpha}^{\alpha \beta} = \epsilon_{\dot \alpha \dot \beta} \epsilon^{\gamma [ \alpha} \beta_{\gamma}^{\beta] \dot \beta},
\eeq

\item $\mathcal{N}=(3,2)$ \\
For the BPS configurations with $\mathcal N = (3,2)$ preserved supersymmetry,
the corresponding BPS equations have
$SU(2)$ symmetry for which both $Y^\alpha~(\alpha=1,3)$
and $Y^{\dot \alpha}~(\dot \alpha = 2,4)$ are in $\mathbf 2$
and $SO(2)$ symmetry which rotates these two doublets.
The BPS equations are given by
\beq
D_2 Y^\alpha = 2 \beta^{\alpha \beta}_{\beta}, \hs{10} D_0 Y^\alpha = D_1 Y^\alpha = 0, \\
D_2 Y^{\dot \alpha} = 2 \beta^{\dot \alpha \dot \beta}_{\dot \beta}, \hs{10} D_0 Y^{\dot \alpha} = D_1 Y^{\dot \alpha} = 0.\,
\eeq
\beq
\beta_\alpha^{\dot \beta \beta} = \frac{1}{2} \delta_\alpha^\beta (\beta_\gamma^{\dot \beta \gamma} + \beta_{\dot \gamma}^{\dot \beta \dot \gamma} ) - \delta_\alpha^{\dot \beta-1} \beta_{\dot \gamma}^{\beta+1, \dot \gamma}, \\
\beta_{\dot \alpha}^{\beta \dot \beta} = \frac{1}{2} \delta_{\dot \alpha}^{\dot \beta} (\beta_{\dot \gamma}^{\beta \dot \gamma} + \beta_{\gamma}^{\beta \gamma} ) - \delta_{\dot \alpha}^{\beta+1} \beta_{\gamma}^{\dot \beta-1, \gamma},\,
\eeq
\beq
\epsilon_{\dot \alpha \dot \beta} \beta_\alpha^{\dot \alpha \dot \beta} = \epsilon_{\alpha \beta} (\beta_\gamma^{\beta \gamma} + \beta_{\dot \gamma}^{\beta \dot \gamma}), \hs{10}
\epsilon_{\alpha \beta} \beta_{\dot \alpha}^{\alpha \beta} = \epsilon_{\dot \alpha \dot \beta} (\beta_\gamma^{\dot \beta \gamma} + \beta_{\dot \gamma}^{\dot \beta \dot \gamma}).
\eeq

\end{itemize}

\paragraph{$1/2$ BPS equations}
\begin{itemize}
\item $\mathcal{N}=(3,3)$ \\
For the BPS configurations with $\mathcal N = (3,3)$ preserved supersymmetry,
the symmetry of the corresponding BPS equations is $SO(4)$ defined by
\beq
U^T g \, U = g, \hs{10} U \in SU(4)_R,
\eeq
where the $SO(4)$ invariant tensor $g$ is given by
\beq
g \equiv \ba{c|c} & i \sigma_2 \\ \hline -i \sigma_2 & \ea.
\eeq
The scalar fields $Y^A~(A=1,2,3,4)$ are in $\mathbf 4$ of $SO(4)$.
The BPS equations are given by
\beq
D_2 Y^A = - \frac{1}{3} \epsilon^{ABCD} \beta_{BCD}, \hs{10} D_0 Y^A = D_1 Y^A = 0.
\eeq
\beq
\beta_{ABC} = \beta_{[A B C]},
\eeq
where we have lowered the indices by using the $SO(4)$ invariant tensor $g$ as
\beq
\beta_{ABC} \equiv g_{B B'} g_{CC'} \beta_A^{B'C'}.
\eeq

\item $\mathcal{N}=(4,2)$ \\
For the BPS configurations with $\mathcal N = (4,2)$ preserved supersymmetry,
the symmetry of the corresponding BPS equations is
$SU(2) \times SU(2) \times U(1)$ for which
$Y^\alpha~(\alpha =1,3)$ and $Y^{\dot \alpha}~(\dot \alpha = 2,4)$ are
in $(\mathbf 2, \mathbf 1)_1$ and $(\mathbf 1, \mathbf 2)_{-1}$, respectively.
The BPS equations are given by
\begin{eqnarray}
D_{2} Y^{\alpha} &=& \beta^{\alpha \beta}_{\beta}, \hs{5} D_0 Y^\alpha = D_1 Y^\alpha = 0, \\
D_{2} Y^{\dot{\alpha}}
 &=& \beta^{\dot \alpha \dot{\beta}}_{\dot{\beta}}, \hs{5} D_0 Y^{\dot \alpha} = D_1 Y^{\dot \alpha} = 0,
\end{eqnarray}
\beq
\beta^{\beta \dot{\alpha}}_\alpha = \beta^{\dot{\beta} \alpha}_{\dot{\alpha} }
 = \beta^{\alpha \beta}_{\dot{\alpha} } = \beta^{\dot{\alpha} \dot{\beta}}_{\alpha} = 0.
\eeq

\item $\mathcal{N}=(5,1)$ \\
The matrix $\Xi$ for the 1/12 BPS configurations
is invariant under $SO(5)$,
so that the corresponding BPS equations have
$Spin(5) \cong USp(4)$ symmetry defined by
\beq
U^T \Gamma_6 U = \Gamma_6, \hs{10} U \in SU(4).
\eeq
The scalar fields $Y^A~(A=1,2,3,4)$ are in $\mathbf 4$ of $USp(4)$.
The BPS equations are given by
\beq
D_2 Y^A = \frac{1}{2} \beta_D^{AD}, \hs{10} D_0 Y^A = D_1 Y^A = 0.
\eeq
\beq
\beta_A^{BC} = \frac{1}{4} \left[ (\Gamma_6^\dagger)^{BC} (\Gamma_6)_{AD} - 2 \delta_A^{[B} \delta_D^{C]} \right] \beta_E^{DE} .
\eeq

\item $\mathcal{N}=(6,0)$ \\
Since the matrix $\Xi$ for $\mathcal N = (6,0)$ BPS configurations
is invariant under $SO(6)_R$, the corresponding BPS equations have
$SU(4)_R$ symmetry for which the scalar fields $Y^A~(A=1,2,3,4)$ are
in the fundamental representation. The BPS equations are given by
\beq
D_{2} Y^{A} = \frac{1}{3} \beta^{AB}_{B}, \hs{10} (D_0 - D_1) Y^A = 0.
\eeq
\beq
\beta^{BC}_{A} = -\frac{2}{3} \delta_{A}^{[B} \beta_{D}^{C]D} .
\eeq
\end{itemize}

\subsection{M2-branes and M5-branes}
For the BPS configurations with M2-branes and M5-branes,
the preserved supercharges are specified by the conditions of the form
\beq
\gamma_2 \Xi_{ij} \epsilon_j = \epsilon_i, \hs{10}
\gamma_1 \Xi'_{ij} \epsilon_j = \epsilon_i,
\eeq
where $\Xi$ and $\Xi'$ are 6-by-6 real symmetric matrices
satisfying $\Xi^2=\Xi'^2=\mathbf 1_6$.
The BPS equations below preserve $n~(n=1,2,3)$ supercharges
and specified by the following matrices
\beq
\Xi = \diag ( \sigma_3, \ast_2 ,\ast_2), \hs{5}
\Xi' = \diag ( \sigma_1, \ast_2 , \ast_2 ) \hs{10} \mbox{for}~n=1, \\
\Xi = \diag ( \sigma_3, \sigma_3 , \ast_2 ), \hs{5}
\Xi' = \diag ( \sigma_1, \sigma_1 , \ast_2 ) \hs{10} \mbox{for}~n=2, \\
\Xi = \diag ( \sigma_3, \sigma_3 , \sigma_3), \hs{5}
\Xi' = \diag ( \sigma_1, \sigma_1 , \sigma_1 ) \hs{10} \mbox{for}~n=3.
\eeq
The explicit form of the BPS equations is given as follows,
\beq
-D_0 Y^B (\Gamma_j)_{BA} (\Xi \Xi')_{ji} + D_1 Y^B (\Gamma_j)_{BA} \Xi'_{ji} + D_2 Y^B (\Gamma_j)_{BA} \Xi_{ji} + \Upsilon_A^{BC} (\Gamma_i)_{BC} = 0,
\eeq
where $1 \leq i,j \leq 2n$.
The BPS bound for energy is given by
\beq
E &\geq& \phantom{+} \frac{k}{4\pi n} \int d^2 x \, \epsilon^{MN} \p_M \tr \left[ Y_A^\dagger D_N Y^B \right](\Gamma_i \Gamma_j^\dagger )_B{}^A (\Xi \Xi')_{ji} \notag \\
&{}& + \frac{k}{8\pi n} \int d^2 x \, \p_1 \tr \left[ Y_A^\dagger \Upsilon_D^{BC} \right] (\Gamma_i)_{BC} (\Gamma_j^\dagger)^{AD} \Xi'_{ij} \notag \\
&{}& + \frac{k}{8\pi n} \int d^2 x \, \p_2 \tr \left[ Y_A^\dagger \Upsilon_D^{BC} \right] (\Gamma_i)_{BC} (\Gamma_j^\dagger)^{AD} \Xi_{ij},
\eeq
and reduces to the following form for the BPS configurations
\beq
E &=& \phantom{+} \frac{k}{8\pi n} \int d^2 x \, \epsilon^{MN} \p_M \tr \left[ Y_A^\dagger D_N Y^B \right](\Gamma_i \Gamma_j^\dagger )_B{}^A (\Xi\Xi')_{ji} \notag \\
&{}& + \frac{k}{4\pi} \int d^2 x \, \p_M \tr \left[ Y_A^\dagger D^M Y^A \right].
\eeq
\begin{itemize}
\item 1/12 BPS equations \\
For the 1/12 BPS configurations with matrices
\renewcommand\arraystretch{1.5}
\beq
\Xi = \ba{c|c}
\sigma_3 & \\ \hline
& \phantom{(} \mbox{{\Huge $\ast$}}_4 \ea, \hs{10}
\Xi' = \ba{c|c}
\sigma_1 & \\ \hline
& \phantom{(} \mbox{{\Huge $\ast$}}_4 \ea,
\eeq
the symmetry of the BPS equations is $SU(2) \times SU(2)$ for which
$Y^a~(a=1,2)$ and $Y^{\dot a}~(\dot a = 3,4)$ are
in $(\mathbf 2, \mathbf 1)$ and $(\mathbf 1, \mathbf 2)$, respectively.
The BPS equations are given by
\renewcommand\arraystretch{1}
\beq
D_0 Y^a = - i ( \beta_d^{ad} - \beta_{\dot d}^{a \dot d} ), \hs{10} (D_1 + i D_2) Y^a = -i \epsilon^{a d} \epsilon_{\dot b \dot c} \beta_d^{\dot b \dot c}, \\
D_0 Y^{\dot a} = - i ( \beta_d^{\dot ad} - \beta_{\dot d}^{\dot a \dot d} ), \hs{10} (D_1 - i D_2) Y^{\dot a} = \phantom{-}i \epsilon^{\dot a \dot d} \epsilon_{b c} \beta_{\dot d}^{bc}.\,
\eeq

\item 1/6 BPS equations \\
For the 1/6 BPS configurations with matrices
\beq
\Xi = \ba{c|c|c}
\sigma_3 & & \\ \hline
& \sigma_3 & \\ \hline
& & \mbox{{\Large $\ast$}}_2 \ea, \hs{10}
\Xi' = \ba{c|c|c}
\sigma_1 & & \\ \hline
& \sigma_1 & \\ \hline
& & \mbox{{\Large $\ast$}}_2 \ea,
\eeq
the BPS equations have $U(1)^2$ symmetry, for one of which
$Y^\alpha~(\alpha=1,3)$ and $Y^{\dot \alpha}~(\dot \alpha=2,4)$
have plus and minus charges.
The other $U(1)$ transformation is defined by
\beq
Y^1 \rightarrow \cos \theta \, Y^1 + i \sin \theta \, Y^3, \hs{10}
Y^3 \rightarrow i \sin \theta \, Y^1 + \cos \theta \, Y^3.
\eeq
The BPS equations are given by
\beq
D_0 Y^a = - i ( \beta_d^{ad} - \beta_{\dot d}^{a \dot d} ), \hs{10} (D_1 + i D_2) Y^a = -i \epsilon^{a d} \epsilon_{\dot b \dot c} \beta_d^{\dot b \dot c}, \\
D_0 Y^{\dot a} = - i ( \beta_d^{\dot ad} - \beta_{\dot d}^{\dot a \dot d} ), \hs{10} (D_1 - i D_2) Y^{\dot a} = \phantom{-}i \epsilon^{\dot a \dot d} \epsilon_{b c} \beta_{\dot d}^{bc}.\,
\eeq
\beq
(D_1-iD_2) Y^1 = 2i \beta_4^{32}, \hs{10} (D_1+iD_2) Y^3 = 2i \beta_2^{41},
\eeq
\beq
\beta_1^{A1} = \beta_3^{A3}, \hs{10} \beta_1^{A3} = \beta_3^{A1}.
\eeq

\item 1/4 BPS equations \\
For the 1/4 BPS configurations with matrices
\beq
\Xi = \ba{c|c|c}
\sigma_3 & & \\ \hline
& \sigma_3 & \\ \hline
& & \sigma_3 \ea, \hs{10}
\Xi' = \ba{c|c|c}
\sigma_1 & & \\ \hline
& \sigma_1 & \\ \hline
& & \sigma_1 \ea,
\eeq
the symmetry of the BPS equations is $SO(3)$ defined by
\beq
U^T g \, U = g, \hs{10} U \in SU(3), \hs{10} g \equiv \diag (-1\,,\, 1\,,\,1\,).
\eeq
The scalar fields $Y^i~(i=1,2,3)$ and $Y^4$ are
in $\mathbf 3$ and $\mathbf 1$ of $SO(3)$ symmetry.
The BPS equations are given by
\beq
D_0 Y^i = i \beta_4^{i4}, \hs{7} (D_1+iD_2) Y^i = \frac{i}{3} \epsilon^{ijk} \beta_{j4k}, \hs{7} (D_1-iD_2) Y^i = -i \epsilon^{ijk} \beta_{4jk},
\eeq
\beq
D_0 Y^4 = - \frac{i}{3} \beta_i^{4i}, \hs{10} (D_1 - i D_2) Y^4 = \frac{i}{3} \epsilon^{ijk} \beta_{ijk}.
\eeq
\beq
\beta_{ijk} = \beta_{[ijk]}, \hs{10} \beta_{i4j} + \beta_{j4i} = \frac{2}{3} g_{ij} \beta_k^{4k}.
\eeq
Here, we have used the $SO(3)$ invariant tensor $g$
to lower the indices
\beq
\beta_{ijk} = g_{jj'} g_{kk'} \beta_i^{j'k'}, \hs{10} \beta_{i4j} \equiv g_{jj'} \beta_i^{4j}.
\eeq
\end{itemize}

\end{appendix}

\end{document}